\documentclass[aps,pre,twocolumn,showpacs,floatfix,superscriptaddress]{revtex4-2}
\usepackage{graphicx}
\usepackage{amssymb,bm}
\usepackage{color}
\usepackage{amsmath}
\usepackage{ulem}
\usepackage{mathrsfs}

\begin{document}

\title{Defects and their Time Scales in Quantum and Classical Annealing \\ of the Two-Dimensional Ising Model}
  
\author{Phillip Weinberg}
\altaffiliation[Present address: ]{QuEra, Cambridge, Massachusetts, USA}
\affiliation{Department of Physics, Boston University, 590 Commonwealth Avenue, Boston, Massachusetts 02215, USA}
\affiliation{Department of Physics, Northeastern University, Boston, Massachusetts 02115, USA}

\author{Na Xu}
\altaffiliation[Present address: ]{Amazon, Seattle, Washington, USA}
\affiliation{Department of Physics, Boston University, 590 Commonwealth Avenue, Boston, Massachusetts 02215, USA}

\author{Anders W. Sandvik}
\email{Communicating author: sandvik@bu.edu}
\affiliation{Department of Physics, Boston University, 590 Commonwealth Avenue, Boston, Massachusetts 02215, USA}
\affiliation{School of Physical and Mathematical Sciences, Nanyang Technological University, Singapore}

\begin{abstract}
  We investigate system-size and ramp-velocity scaling of defects remaining in the two-dimensional transverse-field Ising ferromagnet on
  periodic $L\times L$ lattices after quantum annealing from high to vanishing field. Though the system sizes that we can study using exact integration
  of the Schr\"odinger equation are limited to $L \le 6$, we clearly observe the critical Kibble-Zurek (KZ) time scale $\propto L^{z+1/\nu}$
  (with 3D Ising exponents, $z=1$ and $\nu \approx 0.63$) at the quantum phase transition. We also observe KZ scaling of the gap-protected ground-state
  fidelity at the end of the process. Other quantities in the ordered phase depend on excited states evolving by coarsening dynamics of confined defects,
  with a time scale $\propto L^2$, and interface fluctuations of system-spanning stripe defects with life time scaling as $L^3$. Some of our methods
  and conclusions build on analogies with classical simulated annealing, for which we also obtain new insights. We characterize
  in detail the evolution of system-spanning domains---topological defects characterized by winding numbers in periodic systems---and find differences
  in the dynamic scales of domains with winding numbers $W=(1,0)/(0,1)$ (those spanning the system only horizontally or vertically) and $W=(1,1)$
  (spanning the system diagonally). The former decay on a time scale $\propto L^3$ in the ordered state at fixed temperature while the latter exhibits
  a longer scale $\propto L^{3.4}$. As a consequence of $L^{3.4}$ exceeding the classical KZ scale $L^{z+1/\nu}$ with $\nu=1$ and $z\approx 2.17$, the
  probability of the system hosting $W=(1,1)$ domains scales with the KZ exponent even when observed in the final $T=0$ state, where conventional
  observables are governed by the $L^2$ and $L^3$ ordering time scales. We observe the same time scales in systems with open boundary conditions, except for
  the lack of the slower time scale of diagonal domain walls, which in this case smoothly evolve to horizontal or vertical walls.
  In QA, the $L^3$ time scale for elimination of $W=(1,0)/(0,1)$ domains through interface
  fluctuations also exceeds the relevant KZ scale and the KZ mechanism is therefore responsible for elimination of these domains. The $L^3$ scale can
  nevertheless be detected in the collective evolution of the excited states, using a method of analysis that we develop for this purpose and which
  should also be applicable in QA experiments. The presence of three different time scales at the end of the protocol offers opportunities for
  rigorous multifaceted tests of QA devices with the methods developed here.
\end{abstract}

\date{\today}

\maketitle

\section{Introduction}

The metallurgical method of annealing a crystalline solid to reduce its density of defects
has lent its name to the metaheuristic optimization algorithms of simulated annealing (SA)
\cite{kirkpatrick83,cerny85,granville94} and quantum annealing (QA) \cite{finnila94,kadowaki98,brooke96,brooke99,santoro02}. In SA, a Monte Carlo (MC)
simulation of a model whose ground state represents a hard optimization problem, e.g., the paradigmatic Ising spin glass, is performed with a
gradually lowered temperature, eventually freezing the system into a low-energy state \cite{edwards75,barahona82,kirkpatrick83}. In contrast to this
classical computational scheme, QA is envisioned as a potentially more powerful optimization method when implemented as a physical device, in which
quantum fluctuations of coupled qubits are regulated at very low temperature with the aim of reaching the same classical ground states as in SA.
Promising programmable devices suitable for QA experiments include arrays of superconducting qubits
\cite{johnson10,harris18,weinberg20,zhou21,king22,ali24,king23,king25,sathe25,andersen24} and Rydberg atoms \cite{keesling19,scholl20,ebadi22,manovitz25,zhang25}.
Our work reported here is in a large part motivated by the need to quantitatively understand the ground-state convergence of QA processes in these
systems. To this end, we will also study SA, which in some respects provides a useful analogy to QA though also differing from QA in crucial ways that
we will elucidate here.

A main focus on experimental QA has been to observe Kibble-Zurek (KZ) \cite{kibble76,zurek85} scaling---the build-up of critical fluctuations in
the neighborhood of a continuous quantum phase transition \cite{polkovnikov05,zurek05,dziar05} as the annealing time is increased. For an
annealing process starting from a disordered state and ending in an ordered phase (conventional, ``glassy'', or topological) a phase transition must
be traversed. In the case of a transition that is continuous in equilibrium, the interplay of the KZ mechanism and other ordering processes
\cite{bray94,blanchard17} poses a challenging out-of-equilibrium many-body problem, in QA as well as in SA
\cite{biroli10,chandran12,chandran13a,chandran13b,gagel15,samajdar24} (and similar phenomena arise also when annealing through a weak first-order
transition \cite{pelissetto17,suzuki24}). Given the increasing experimental attention and capability to detect production
of different defects and their time scales \cite{gao22,ali24,manovitz25,zhang25}, further theoretical and computational studies of nonequilibrium
criticality and ordering kinetics are called for. We here consider the prototypical example of the two-dimensional (2D) Ising model by numerical SA and QA,
systematically investigating the ordering dynamics in terms of both the growth of critical fluctuations close to the phase transition and the subsequent
formation of long-range order through other mechanisms.

\begin{figure}[t]
\includegraphics[width=70mm]{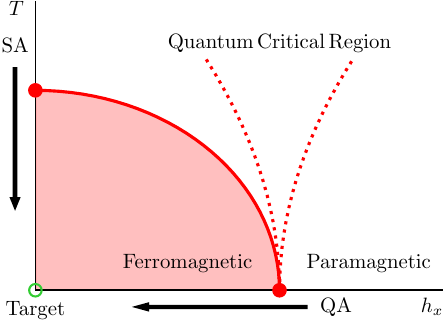}
\caption{Phase diagram of the 2D TFIM in the plane of temperature ($T$) and transverse field ($\Gamma$).
The two arrows represent the annealing paths taken here in QA and SA processes. The green circle indicates the target classical ground state.  
While the quantum-critical $T>0$ regime is not traversed with the paths taken here, a finite QA velocity has effects similar to $T>0$, with
energy pumped into the system in the neighborhood of the critical point and the excitations subsequently thermalizing.}
\label{fig:phase_diag}
\end{figure}

Figure \ref{fig:phase_diag} shows a schematic phase diagram of the 2D $S=1/2$ transverse-field Ising model (TFIM) that we will study under
both QA and SA. The Hamiltonian is
\begin{equation}\label{qham}
H = -J\sum_{\langle ij\rangle} \sigma_i^z\sigma_j^z - \Gamma \sum_{i=1}^N \sigma_i^x,
\end{equation}  
where $\sigma^z_i$ and $\sigma^x_i$ are the standard Pauli matrices, $\langle ij\rangle$ refers to nearest neighbors on the 2D square lattice, and
$J,\Gamma \ge 0$. In principle, any path in the plane of temperature and coupling ratio $\Gamma/J$ can be taken when annealing to the classical
ground state at $T=0$, $\Gamma=0$. Here we will consider the two limiting cases of pure SA and pure QA, as illustrated in Fig.~\ref{fig:phase_diag}.
In SA we set the field strength to $\Gamma=0$ and use MC simulations with local single-spin updates to anneal the classical system versus the temperature
$T$ from far above the transition temperature to $T=0$. In QA, using exact
numerical integration of the Schr\"odinger equation for small systems, we start from the ground state at $J=0,\Gamma=1$ and change the couplings versus
time such that $J\to 1$ and $\Gamma \to 0$ at the end of the process. In either case, the equilibrium phase transition (emerging in the limit of infinite
system size) is continuous and  critical fluctuations can be observed in the neighborhood of the transition temperature $T_c$ (in SA) or coupling ratio
$(\Gamma/J)_c$ (in QA) even if the process is not completely adiabatic (in QA) or quasi-static (in SA). However, the associated KZ scaling is only one of
the interesting aspects of the annealing processes, with the subsequent stage of reaching deep into the ordered phase presenting additional complexity.

Though the Ising ferromagnetic ground state is trivial, there are still open issues regarding the efficiency of QA and SA processes to reach it as a
function of the system size and the annealing time. Even though our numerical simulations of QA reach only up to $6\times 6$ spins, we still observe
remarkably good scaling behaviors with no less than three distinct time scales as $\Gamma\to 0$, related to criticality (the KZ mechanism) as well as
the kinetic processes leading to the ordered phase by coarsening of confined domains and slow fluctuations of system-spanning domain walls. Many of our
methods of data analysis and interpretations of results for QA are guided by analogies with SA, which we study first using large lattices ($L$ up to
several hundred). In particular, we obtain
new qualitative and quantitative insights into the development and persistence of system-spanning topological defects. This aspect of our work also relates
more broadly to generic questions about the stability of large-scale defects in quantum and classical many-body systems out of equilibrium, with the
2D Ising models serving as paradigmatic examples for which reliable numerical results can be obtained---in the quantum case perhaps surprisingly, given
that we only study systems with up to $36$ spins.

In QA, we study not only physical observables in terms of their conventional expectation values, but also develop a scheme to isolate the evolution of the
excited states from the dominant ground state at long annealing times---a method that can also be implemented experimentally. Our work firmly
establishes the 2D Ising model as a non-trivial system for multifaceted tests of experimental QA devices. Moreover, we propose specific QA experiment
that may provide further insights into emergent collective ordering kinetics beyond what can be derived from the small systems studied here.
Our study can also serve as a benchmark for more sophisticated but approximative numerical solutions of the Schr\"odinger equation, which so far have
not been able to study ordering kinetics of the 2D TFIM and related models at the level achieved here. Though much larger systems were simulated in
some of the past studies \cite{schmitt22,king25,tindall25,mauron25}, they were limited by the break-down of assumptions and accumulation of integration errors to
very short annealing times, corresponding to only to the initial stages of ordering.

Below we give a brief synopsis the main issues addressed and the key results obtained. The KZ scaling and defect ``evaporation''
aspects of the study are summarized in Secs.~\ref{sec:KZ} and \ref{sec:final}, respectively, and the organization of the rest of the paper is 
outlined in Sec.~\ref{sec:outline}.

\subsection{Kibble-Zurek scaling}
\label{sec:KZ}

The first interesting stage of a QA or SA process is when the control parameter traverses its critical value. The way the finite annealing rate affects
critical scaling properties can be understood in terms of an ansatz originating in early works by Kibble \cite{kibble76} and Zurek \cite{zurek85} on
topological defects. The originally classical KZ ideas were later generalized to quantum systems \cite{polkovnikov05,zurek05,dziar05} and further
developed for different annealing protocols and methods of analysis
\cite{zhong05,dziar10,polk11,degrandi10,degrandi11,chandran12,chandran13a,chandran13b,degrandi13,liu13,delcampo14,liu14,liu15,rubin17,xu17,xu18}.
KZ scaling behaviors
follow from the relationship between the correlation length $\xi$ and the relaxation time $\tau$, which at a conventional classical or quantum phase
transition is $\tau \propto \xi^z$, where $z$ is the dynamic exponent and $\xi$ diverges as $\delta^{-\nu}$ when the distance $\delta$ to a critical point
is taken to zero; in our cases, Fig.~\ref{fig:phase_diag}, $\delta=T-T_c$ when $\Gamma=0$ and $\delta=\Gamma-\Gamma_c$ when $T=0$ (and we neglect any
minus sign, treating $\delta$ as $|\delta|$). In terms of the rate (velocity) $v$ by which the relevant parameter is changed when passing through
the critical point, the divergence of the correlation length is cut off at a velocity dependent maximum value,
\begin{equation}\label{xiv}
\xi_v \propto v^{-\nu/(z\nu+1)},
\end{equation}  
which is attained when the annealing parameter is close to its critical value and $v$ is sufficiently small.

In QA, $z$ is the intrinsic dynamic exponent, i.e., $z=1$ in the TFIM, while in SA it corresponds to the stochastic process applied, e.g., $z \approx 2.17$
for local MC updates (e.g., Glauber dynamics with single-spin flips used here) in the classical 2D Ising model \cite{ngale00,liu14}. Velocity-limited critical
forms of physical observables $A(\delta)$ in the thermodynamic limit can be obtained from the conventional equilibrium critical expressions by using
$\delta \propto \xi^{-1/\nu}$ and then replacing the equilibrium correlation length $\xi$ by its velocity limited form $\xi_v$ in Eq.~(\ref{xiv}).

The KZ mechanism also implies a power-law relationship between the quasi-static or -adiabatic annealing time and the system size, which provides
a useful way to analyze numerical simulation results as well as QA experiments. This is the form in which we will consider KZ scaling
here. From Eq.~(\ref{xiv}), by replacing $\xi_v$ by the system length $L$, the KZ time scale that has to be attained for the system to remain in
equilibrium all the way to the infinite-size critical point ($\delta=0$) is
\begin{equation}
t_{\rm KZ} \propto L^{z+1/\nu}.
\label{taukz}
\end{equation}  
For a generic quantity $A$ with equilibrium critical form $A \propto \delta^\kappa$, with a corresponding generic exponent $\kappa$,
the velocity and size dependence when the annealing has reached the critical point can be written as
\begin{equation}
A(v,L)= L^{-\kappa/\nu} f(vL^{z+1/\nu}),
\label{chiv}
\end{equation}  
where the analytic scaling function $f(x)$ takes a constant value for $x\to 0$ and $L$ has to be sufficiently large for finite-size scaling corrections
to be negligible. For large $x$, the trivial size dependence of $A$ when $\xi_v \ll L$, e.g., $L$ independent or $A \propto L^{-d}$, demands a specific
power-law form of $f(x)\sim x^a$ \cite{liu14} that is easily obtained from Eq.~(\ref{chiv}) by choosing the exponent $a$ such that
$L^{-\kappa/\nu} (vL^{z+1/\nu})^a$ contains the appropriate power of $L$, thus also producing the velocity dependence.

Our main aim here is to study kinetic processes affecting QA or SA beyond the critical point, eventually as $\Gamma \to 0$ or $T \to 0$. In
such a process many different quantities can be computed (or measured in experiments) and they can be analyzed in different ways to reveal scaling behaviors.
We will see that KZ scaling can be observed in some quantities even at the extreme end ($\Gamma = 0$, $T= 0$) of an annealing process, but in different
ways in QA and SA. In QA, the ground state fidelity, i.e., the ``success probability'' in the context of optimization, is protected by the gap in the
ferromagnetic phase and is almost fully locked into its final value as the system passes the quantum-critical point. We indeed observe extremely good
KZ scaling of the ground state fidelity at the final $\Gamma=0$ state of the QA process, despite the system sizes being restricted to $L \le 6$.
Most other physical observables depend in some way on excited states, which are not governed by the KZ mechanism because of the gapless spectrum (for
increasing $L$) above the lowest excited state and the related faster kinetic processes that we will investigate in detail. Likewise, in SA the
system annealed deep into the ordered phase is subject to ordering processes unrelated to the critical fluctuations underlying the KZ mechanism.
Interestingly, we find that the survival time of certain system-spanning stripe defects still obeys KZ scaling, because these defects are absent 
below $T_c$ in equilibrium and their elimination by fluctuations (if imposed by hand in an initial state) is governed by a longer time scale than
the KZ time required for the system to stay in equilibrium when traversing the critical point.

\subsection{Dynamics in the ordered phase}
\label{sec:final}
           
In early discussions of KZ scaling, the assumption was that the topological defects produced when passing through the critical point would stay essentially
frozen when annealing further into an ordered phase. If all dynamical processes in the ordered state are much slower than the KZ rate $v_{\rm KZ}=1/t_{\rm KZ}$,
then the near-critical state with correlation length $\xi_v$ in Eq.~(\ref{xiv}) (assuming $L \gg \xi_v$) remains essentially unchanged all the way to $T=0$
in SA or in the classical limit of the  Hamiltonian in QA. Likewise, a finite system that has stayed in equilibrium all the way to the transition point would
remain close to its critical state also when the process continues into the ordered phase. This aspect of QA is of relevance in experiments when the state can
be accessed only in the classical limit, e.g., $\Gamma\to 0$ in D-Wave devices \cite{johnson10}. Indeed, it was possible to demonstrate KZ scaling in D-Wave
QA of Ising spin glasses thanks to the slow dynamics of the glassy state \cite{king23,king25}. In general, the system of course must always continue to evolve
to some extent also beyond the critical point, and, for sufficiently slow annealing, an ordered state (conventional or glassy) eventually develops through
processes beyond the KZ mechanism, by fluctuations facilitating energy minimization. A finite system will reach its true classical ground state this way
over an extent of time that scales with the system size. In many cases, some of these kinetic ordering processes are actually faster than the KZ scale and
do not rely on a critical state having built up with correlation length $\xi_v \approx L$ first. If all defects can be eliminated faster than the KZ scale,
the critical stage with correlation length $\xi_v \approx L$ can be completely bypassed.

\begin{figure}[t]
\center{\includegraphics[width=60mm,angle=270,clip]{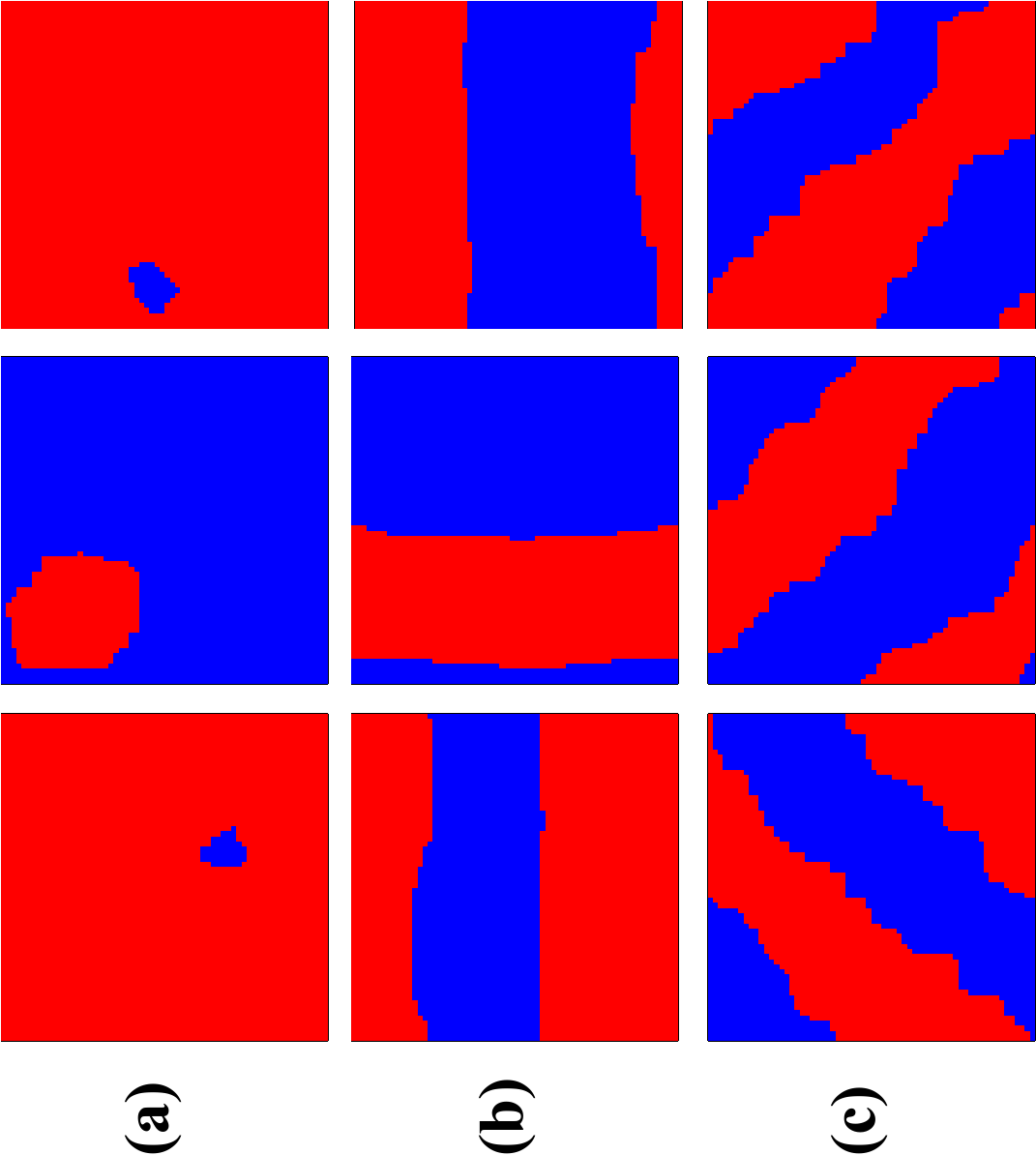}}
\caption{Spin configurations of a classical $L=64$ periodic Ising system (up and down spins shown as red and blue squares)
with defects remaining after annealing from $T=4$, above $T_c \approx 2.27$, to $T=0$ in a linear ramp using a total of $2^{10}$ MC sweeps
(each consisting of $L^2$ random single-spin flip attempts with Metropolis acceptance probability). The three rows exemplify different types
of defects; (a) confined domains with winding number $(0,0)$, which are formed and eventually vanish (for longer annealing times)  through
standard coarsening dynamics, (b) system-spanning stripe domains oriented along a system axis, with winding number $(1,0)$ or $(0,1)$, and (c)
diagonal stripe domains with winding number $(1,1)$. Note that each of the configurations in (b) and (c) only have two domains because
of the periodic boundary conditions. The system spanning domains can vanish through interface fluctuations, when the two domain walls touch each
other, thus breaking the stripe and allowing the resulting defect with winding number $(0,0)$ to shrink as those in (a). The time scale
for eliminating winding domains in SA is longer than the coarsening scale $L^2$.}
\label{fig:config}
\end{figure}

The relevant dynamical processes of course depend on the system under study. In the case of the classical uniform Ising model, coarsening dynamics
\cite{bray94,biroli10} involving confined defects such as those illustrated in Fig.~\ref{fig:config}(a) involves a faster time scale $t_{\rm coa} \propto L^2$
to ordering below $T_c$ than the KZ time $t_{\rm KZ} \propto L^{z+1/\nu}$ required for a finite system to reach its critical state. An annealing time of order
$L^2$ may therefore appear sufficient for the system to reach the ferromagnetic state at $T=0$ from any initial temperature. When reaching the critical point
in such a process, the correlation length is $\xi_v$, with $v \propto L^{-2}$ in Eq.~(\ref{xiv}), which grows slower than $L$ (in both SA and QA). Thus, the
fully critical finite-size state where $\xi \approx L$ is bypassed. Proceeding into the ordered phase, the fractal critical domains of typical
size $\xi_v \ll L$ begin to develop smooth domain boundaries (coarsen) over time. Eventually an ordered state with isolated defects form, as in
Fig.~\ref{fig:config}(a), and the fully ordered state develops when these confined defects shrink away. For $v=\lambda L^{-2}$, with
sufficiently small $\lambda$, all confined domains are eliminated at the end of the anneal ($T=0$, $\Gamma=0$). However, this scenario neglects the
possibility that a time scale longer than $L^2$ may be required to remove topological, system-spanning domains that appear in some annealing instances
\cite{biroli10} and are illustrated in Figs.~\ref{fig:config}(b) and \ref{fig:config}(c). As far as we are aware, the processes of final elimination of
these system-spanning defects have not been quantitatively studied, neither in QA nor SA of the 2D Ising model, though similar large-scale defects are at
the heart of the seminal KZ works \cite{kibble76,zurek85}.

System-spanning defects in the classical 2D Ising model have been studied in other contexts.
In instantaneous MC quenches from $T=\infty$ to $T=0$ \cite{spirin01a,spirin01b,barros09,olejarz12} it was observed that a finite
fraction of final states remain with completely straight domain walls (which cannot be removed by single-spin flips at $T=0$) while diagonally
oriented domains can shrink away and do so at a time scale $L^a$, with the value of the exponent estimated at $a \approx 3.5$. It was also argued that
horizontally or vertically oriented domains vanish on a shorter time scale $L^3$ at low temperature \cite{spirin01a,lipowski99}. The apparently different
behaviors of these two types of system-spanning domains at $T>0$ has not been explained quantitatively, though related work suggests non-universal
interface diffusivity \cite{plischke87}. Here we will investigate the stochastic-dynamical aspects of system spanning domain walls in SA and obtain
further insights into their relevant time scales and how they can be observed. We also carry out extensive studies of the time scales of elimination
of horizontal and diagonal system-spanning domains by domain-wall fluctuations at fixed temperatures below $T_c$. We confirm the $L^3$ time scale for
elimination of domain walls in the case of $W=(1,0)$ and obtain a more precise value of the exponent for $(1,1)$ domains; $a=3.42 \pm 0.02$.

The $W=(1,1)$ diagonal domains are particularly interesting, as their probability of survival in an SA process crossing the critical temperature obeys KZ
scaling even in the final $T\to 0$ state, because of their longer time scale of vanishing through domain-wall diffusion below $T_c$. While diagonal domains
are present in equilibrium only close to $T_c$ (in a window of size scaling as $L^{-1/\nu}$, as we will also show explicitly) in SA they are present also
below $T_c$ with a probability set by KZ scaling, with their decay taking place primarily very close to $T_c$. Their overall impact on most physical
observables is minor, however, and only the $L^2$ and $L^3$ scales are clearly visible when $T \to 0$ in observables such as the order parameter and the
excess energy. To observe the KZ scaling of diagonal domains in the ordered phase, the system has to be probed explicitly for this kind of domains, using
the corresponding winding-number probability (in periodic systems) or other quantities that vanish when $T \to 0$ unless a diagonal domain is present.

We also investigate system-spanning domain walls on lattices with open boundary conditions, for which the winding number cannot be defined. Indeed, in
this case we find that an imposed diagonal domain wall (between opposite corners)
decays away on the coarsening time scale $L^2$, either resulting in a fully ordered state or, as
an intermediate step (the final step if $T=0$), first evolving into a horizontal or vertical domain wall. In the latter case, we find a logarithmic (log)
correction to the $L^2$ scaling, and the time to final elimination of the remaining system-spanning domain wall is still of order $L^3$, which is reflected
in physical observables in the same way as in the periodic systems.

In QA of the TFIM, the defects and their time scales inside the ordered state have been less studied than their classical counterparts. There is evidence to
suggest that QA of a system described by a generic non-integrable Hamiltonian is governed by classical hydrodynamic behavior deep inside an ordered phase,
as a consequence of the system thermalizing with the excess energy acquired when passing the phase transition \cite{chandran13a,shimizu18,libal20}. A competing
theory asserts that the defect production is determined simply by Landau-Zener (LZ) transitions
between a few low energy eigenstates near the critical point \cite{zurek05,mitchell05,caneva08,zanca16,wauters17,zeng23,grabarits25}. To our knowledge,
the LZ scaling argument has only been tested on integrable 1D models or in mean-field theories, however. The correct description of the defect dynamics in
the ordered phase in integrable models follows from an extensive number of conserved quantities, which leads to the break-down of classical hydrodynamics.
Regardless of details, these dynamic processes may again possibly be faster than the critical KZ dynamics. In the 1D TFIM, the KZ and LZ exponents are in
fact identical, which makes it more difficult (though not impossible) to distinguish the two mechanisms \cite{king22,grabarits25}.

In the specific case of the
2D TFIM considered here, a recent numerical study based on time evolved matrix- and tensor-product states \cite{schmitt22} was not able to reach far
inside the ordered phase, but a generalized KZ mechanism was argued to describe the available near-critical data, and signs of quantum coarsening were
observed further inside the ordered phase. However, the coarsening time scale was not determined.
Though recent progress has been made on more efficient numerical algorithms \cite{tindall25,mauron25}, in general it is still very difficult to solve
the Schr\"odinger equation on size and time scale now accessible in 2D and 3D QA experiments \cite{king23,king25,manovitz25}.

In our exact numerical solutions of TFIMs with $L \le 6$, there is no limitation on the duration of the QA process. Even though the systems are very
small compared to those used in the classical SA studies, and even in the recent study of QA with matrix- and tensor-product states \cite{schmitt22},
we find that QA is perfectly consistent with coarsening dynamics with the same time scale as in the classical SA case; $t_{\rm coa} \propto L^2$.
This time is again shorter than the KZ time scale $t_{\rm KZ} \propto L^{z+1/\nu}$, now with the 3D Ising exponents
$z=1$ and $\nu \approx 0.63$. Thus, the system can order by coarsening, and KZ scaling can in general not be observed after the annealing parameter
has reached deep into the ordered phase. A notable exception is the ground state fidelity (``success probability''),
which is protected by the gap in the ferromagnetic phase and is almost fully locked into its final value as the system passes the quantum-critical
point. We indeed observe extremely good KZ scaling of the ground state fidelity at the final $\Gamma=0$ state of the QA process. Some perturbed form
of KZ scaling still likely applies very close to the critical point \cite{schmitt22,king23,samajdar24}, but we find that the fast $L^2$ coarsening time
dominates the behavior far inside the ordered phase as the excited states are not gap-protected and appear to develop essentially classical dynamics.
This behavior is consistent with the generic scenario of thermalization of excitations in the ordered phase \cite{chandran13a,shimizu18,libal20}
but has not observed previously with reliable numerics of the prototypical TFIM.

The fate of system-spanning stripe domains has not been addressed specifically in the context of QA (e.g., in Ref.~\onlinecite{schmitt22}), but, because of
the plausible emergence of classical dynamics of a thermalized state, we would expect them to behave much as in SA in the ordered phase. We will
show that $W=(1,0)/(0,1)$ domains can be studied quantitatively even in very small systems and their time scale of elimination in QA is indeed $L^3$, as in the
classical case. However, this time scale is longer than the quantum KZ scale $L^{2.59}$ (unlike SA, in which the KZ scale $L^{3.17}$ exceeds $L^3$) and
its presence in the data is more subtle. We develop a method to study the collective QA evolution of the excited states, where we detect both the $L^2$
and $L^3$ time scales. The system sizes are too small to accommodate proper $W=(1,1)$ domains, and we therefore cannot analyze those specifically in QA.

\subsection{Outline and calculation strategy}
\label{sec:outline}
           
Given that we will use insights from SA to understand novel aspects of QA in the ordered state of the TFIM, we begin in Sec.~\ref{sec:SA} by
exploring the final-stage $T \to 0$ ordering kinetics of the 2D Ising model subject to classical SA. Finite-size KZ scaling in SA of the Ising model has been
extensively studied in the past \cite{liu14} and will only be considered here in our analysis of system-spanning domain walls. In ~Sec.~\ref{sec:sascale} we first
study physical observables, the order parameter and the excess energy, in systems after annealing to $T=0$, detecting the $L^2$ coarsening scale and
the $L^3$ scale of elimination of $W=(1,0)/(0,1)$ stripe domains. In Sec.~\ref{sec:wait} we study the time scales of elimination of $W=(1,0)$ and $W=(1,1)$
domains at fixed $T < T_c$. We return to SA in Sec.~\ref{sec:wind} to quantify how system-spanning domains form and vanish during the SA process, using
winding probabilities as a quantitative measure of the different types of system-spanning defects. All the above simulations are done with periodic
boundary conditions. In Sec.~\ref{sec:open} we investigate systems with open boundaries.

Turning to QA of the TFIM, in Sec.~\ref{sec:QA} we perform exact numerical integration of the Schr\"odinger equation for systems with up to $6\times 6$
spins and use the insights from SA to analyze the results. In Sec.~\ref{sec:QA:nes_end} we first confirm KZ scaling both at the infinite-size quantum-critical
point and, in the case of the fidelity, also at the classical end point of the QA process. In Sec.~\ref{sec:QA:nes_crit} we analyze other physical observables
in the ordered phase, extracting the dominant coarsening time scale $L^2$ as well as the $L^3$ scale we find and associate with the removal of remaining
straight domain walls. In Sec.~\ref{sec:discuss} we conclude with further discussion of our findings and the applicability of the results to QA experiments.

\section{Classical Simulated Annealing}
\label{sec:SA}

We here consider the classical 2D Ising ferromagnet described by the Hamiltonian
\begin{equation}
H_\mathrm{cl} =-J \sum_{\langle  ij\rangle } \sigma_{i} \sigma_{j},~~~ \sigma_i=\pm 1,
\label{ham}
\end{equation}
where the coupling is set to $J=1$ and  $\langle  ij\rangle $ stands for nearest-neighbor sites on the 2D square lattice with $L^2$ spins.
In most cases, in Secs.~\ref{sec:sascale}-\ref{sec:wind}, we apply periodic boundary conditions but in Sec.~\ref{sec:open} we also present results
for systems with open boundaries, to confirm universality of the SA time scales. We also find some important differences related to
system-spanning defects.

We apply SA with single-spin updates (Glauber dynamics) to the system, starting from an equilibrated state at a temperature $T_{\rm init}$ well above
the critical temperature $T_{\rm c}= 2/\ln(1+\sqrt{2}) \approx 2.27$, continuing all the way down to $T=0$ following the standard linear protocol
\begin{equation}
T(t) = T_{\rm init}(1-t/t_{\rm SA}).
\label{eq:tprotocol}
\end{equation}
Here the dimensionless time $t$ is an integer representing the number of MC sweeps, with each sweep involving $N=L^2$ flip attempts of randomly selected
spins subject to the Metropolis acceptance probability. The temperature is changed after each MC sweep. The number of sweeps of the entire SA process
is $t_{\rm SA}$ and we define the annealing velocity $v$ to be the inverse of the total annealing time; $v = 1/t_{\rm SA}$, i.e., not including the
total change of $T$ by $T_{\rm init}$ units.

We have implemented multi-spin coding, where a single 64-bit integer (in some cases 32-bit) represents 64 independent spins; thus each SA
realization contributes 64 distinct contributions to the averages of quantities accumulated at the end of each repetition. The total number of
repetitions depends on the velocity and the system size, from over $10^8$ for small systems to about $10^4$ for the most time-consuming cases of
low $v$ and large $L$. Many of the quantities we study are self-averaging, thus reducing the required number of SA repetitions for large systems.

\subsection{Coarsening and large-scale defects}
\label{sec:sascale}

To study scaling of common physical observables in systems after annealing to $T=0$, we here use lattice sizes from $L=16$ to $L=768$ and velocities
spanning a wide enough range for observing the evolution of the final state from completely disordered to essentially fully polarized. Fig.~\ref{fig:config}
shows some examples of spin configurations at the end of SA runs for an $L=64$ system at velocity $v=2^{-10}$, which is low enough (for this rather small
system) for the final $T=0$ state to typically contain only a small number of defects with respect to the fully polarized ferromagnet (and no defects in a
significant fraction of the instances). The defects can broadly be classified into three different groups represented by the rows of Fig.~\ref{fig:config};
confined domain walls in row (a), system-spanning domain walls winding across the periodic boundaries only in the $x$ or $y$ direction in (b), and diagonal
domain walls winding in both the $x$ and $y$ directions in (c). In these cases each instance hosts a single defect domain, which is typically the case at some
point before a system orders perfectly. In the case of the system-spanning stripe defects, they can also wind around the periodic boundaries multiple times,
beyond the winding once in one lattice direction in Fig.~\ref{fig:config}(b) and once in both directions in Fig.~\ref{fig:config}(c). The higher winding
numbers are naturally less prevalent, and we will present some statistics in Sec.~\ref{sec:wind}.

To quantify the effects of remaining defects, at the end of each annealing instance we first study the mean squared order parameter (magnetization),
\begin{equation}\label{eq:m2}
m^{2} = \left \langle \left ( \frac{1}{N}\sum\limits_{i=1}^N\sigma_{i} \right )^{2} \right \rangle,
\end{equation}
and the Ising energy density (where we use a subscript $z$ for consistency with the later study of the TFIM)
\begin{equation}
e_z=\langle H_\mathrm{cl}\rangle/N,
\end{equation}
which we will analyze relative to the size-independent equilibrium value $e_{z,0}=-2$ for systems with periodic boundary
conditions at $T=0$. 

We will assume $v$ and $L$ scaling forms $L^{-b}f(vL^\alpha)$ for these quantities in analogy with the KZ form Eq.~(\ref{chiv}) with some exponents
$\alpha$ and $b$. We first present simple arguments for the exponent $b$ for $m^2$ and $e_z$: At some point of the process there will be
one or a few large defect domains in an essentially ordered background (possibly with some additional small defects that are not important).
These large domains are initially of size some significant fraction of the system size, i.e., of typical length $l \propto L$ (regardless of
whether these defects are confined or system-spanning). Thus, the deviation of the magnetization density from its maximum value is of order
unity and, therefore, $b=0$ for $m^2$. Close to equilibrium, when $T \to 0$ and $m^2$ approaches $1$, $1-m^2$ should be analyzed with a different
exponent but here we focus on the stage when the large defects are present and gradually shrink versus time. In the case of the energy, the total cost
of the domain walls is then of order $l \propto L$, which for the excess energy density implies $e_z-e_{z,0} \propto L^{-b}$ with $b=1$. Thus, to account
for both size and velocity dependence, we will test the scaling forms
\begin{subequations}
\label{scforms}  
\begin{gather}
m^2(v,L) = f_{m}(vL^{\alpha}), \label{m2scforms} \\
e_z(v,L)=e_{z,0} + \frac{1}{L}f_{e}(vL^{\alpha}), \label{ezscforms}
\end{gather}
\end{subequations}
where the exponent $\alpha$ on the time scale $L^\alpha$ will be obtained from numerical data. As we will see, data in different regimes of $v$ and $L$
exhibit scaling with different values of $\alpha$, i.e., the same SA process can be described by more than one scaling function---at least two in the cases
studied here.

\begin{figure}[t]
\includegraphics[width=7cm, clip]{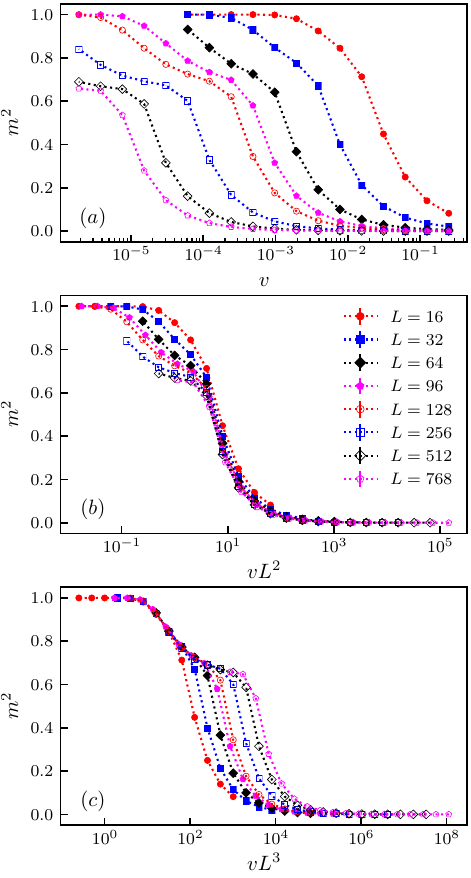}
\vskip-1mm
\caption{Squared magnetization averaged over large numbers of SA instances at $T=0$. (a) Raw data vs the annealing velocity for system sizes
$L=16$, $32$, $64$, $96$, $128$, $256$, $512$, and $768$, according to the legends in panel (b). In (b) and (c), the velocity is rescaled
by $L^2$ and $L^3$, respectively, to demonstrate regions of data collapse with two different time scales in the growth of the order parameter.}
\label{fig:m2all}
\end{figure}

Figure \ref{fig:m2all} summarizes our results for $m^2$. The raw data versus $v$ for several different system sizes are shown in Fig.~\ref{fig:m2all}(a).
In addition to an overall shift of the curves toward lower velocity when the system size increases, for the larger systems there is also a qualitative
change in behavior before the systems approach the fully polarized state with $m^2=1$; the increase is slowed down and a shoulder-like feature forms
at $m^2 \approx 0.6$. As shown in Fig.~\ref{fig:m2all}(b), by graphing the data versus $vL^2$ the overall shift with $L$ when $m^2$ is below the shoulder
value is collapsed onto a common curve for the larger system sizes (for which corrections to the leading behavior are small), thus demonstrating a time
scale $\propto L^2$ and delivering the corresponding scaling function $f_m(vL^2)$ with $\alpha=2$ in Eq.~(\ref{m2scforms}). This is the expected well-known time
scale of coarsening dynamics \cite{bray94}, which was also studied previously in the context of the $T<T_c$ break-down of KZ scaling in SA of the
2D Ising model \cite{biroli10}. Conventional coarsening here refers to the process by which confined domains, such as those in Fig.~\ref{fig:config}(a),
minimize their boundary energy by tending toward circular smooth shapes that also shrink at low temperatures; eventually vanishing in the case of SA to
$T=0$ at sufficiently low $v$. In an infinite system, domains continue to increase in size versus time and the system never reaches the true ground
state \cite{bray94} consisting of a single ferromagnetic domain.

\begin{figure}[t]
\includegraphics[width=7cm,clip]{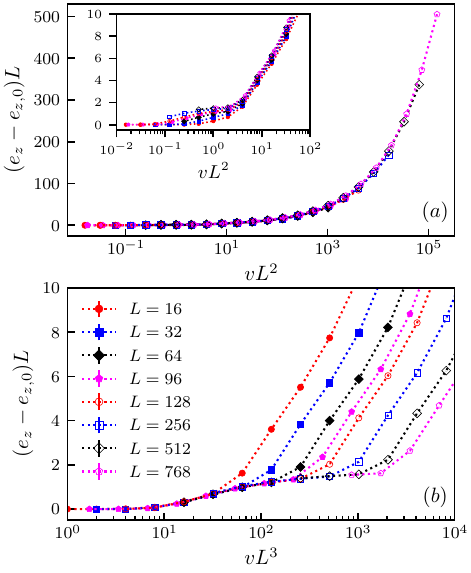}
\vskip-1mm
\caption{Excess mean Ising energy density scaled by the system length $L$ at the end of the same SA runs as those in Fig.~\ref{fig:m2all}.
The data are plotted against $vL^2$ in panel (a) and $vL^3$ in panel (b). The legends in (b) apply to the data sets in (a) as well. The inset of
panel (a) shows a zoomed-in view of the lower left-hand corner of the main panel.}
\label{fig:dE}
\end{figure}

The coarsening time scale does not describe the ultimate convergence to the ground state \cite{biroli10}. For $m^2$ at and above its shoulder
feature, we find a different behavior corresponding to a time scale $\propto L^3$, which we demonstrate by graphing the data versus $vL^3$ in
Fig.~\ref{fig:m2all}(c). Here we observe that the shoulder-like feature extends to larger values of the scaling variable as $L$ increases, likely forming
a plateau at a constant non-zero $m^2$ extending to arbitrarily high $vL^3$ as $L \to \infty$ (as we will argue in more detail below). Thus, some kind
of long-lived defects form in at least a fraction of the SA instances, and it is natural to assume (and we will prove below) that these defects are
the system-spanning stripe domains previously found in fixed-$T$ simulations (with an initial stripe state imposed) \cite{lipowski99} and instantaneous
quenches \cite{spirin01a,spirin01b,olejarz12}.

We observe the same two time scales also in the excess energy density, which is graphed versus $vL^2$ and $vL^3$ in Fig.~\ref{fig:dE}(a) and
Fig.~\ref{fig:dE}(b), respectively. In this case, for the data to collapse onto a common scaling function we also have rescaled the observable by the
system size $L$ in accord with Eq.~(\ref{ezscforms}). In Fig.~\ref{fig:dE}(a) the small-$L$ corrections to the common curve $f_e(vL^2)$ are even smaller
than the more visible corrections to $m^2 \approx 0.6$ in Fig.~\ref{fig:m2all}(b). However, clear deviations from scaling collapse are seen for the smaller
values of $vL^2$ on the more detailed scale in the inset of Fig.~\ref{fig:dE}(a). The data in this range again collapse when graphed versus $vL^3$, as 
shown in Fig.~\ref{fig:dE}(b). Here the formation of a plateau with constant excess energy with increasing $L$ is also apparent and corresponds to the less
clear build-up toward a constant $m^2 \approx 0.6$ in Fig.~\ref{fig:m2all}(c). These plateaus are naturally explainable as consequences of system-spanning
defects remaining stable until $T=0$ without significant change in the mean length of the domain walls, when the confined domains have evaporated away
on the faster $L^2$ time scale. The final low-velocity approach to zero excess energy corresponds to the topological domains finally being broken (after
which they can vanish by conventional coarsening), the time of which follows a broad probability distribution at given $v$, thus giving rise to the smooth
evolution from the plateau value of the average energy to zero in Fig.~\ref{fig:dE}(b). The gradually diminishing probability of a stripe domain surviving
all the way to $T=0$ as $v$ is lowered is likewise manifested as $m^2$ increasing from the plateau value $m^2\approx 0.6$ and finally approaching $1$
in Fig.~\ref{fig:m2all}(c).

\begin{figure}[t]
\includegraphics[width=7.5cm,clip]{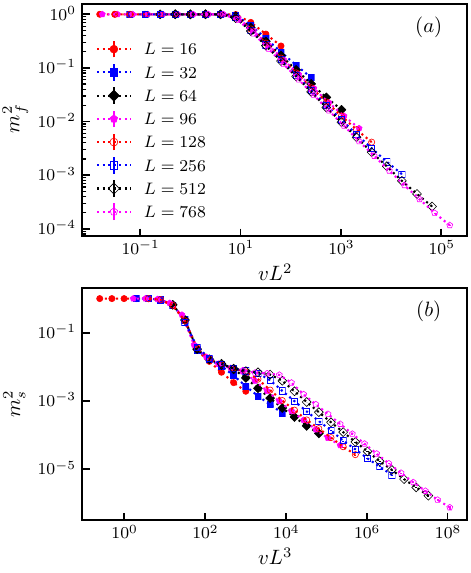}
\vskip-1mm
\caption{Data collapse of the grouped squared order parameters; $m_{\rm f}^2$ in (a) and $m_{\rm s}^2$ in (b), corresponding to averages over the
respective 20\% of the instances classified as fast or slow according to the value of $m^2$. The velocity in (a) and (b) has been rescaled by $L^2$ and
$L^3$, respectively, corresponding to the time scale of conventional coarsening dynamics and the slower removal of stripe defects.}
\label{fig:m2}
\end{figure}

Assuming that the stripe domains are predominantly of the horizontal or vertical kind, i.e., with winding number $(1,0)$ or $(0,1)$, and that only
one such defect domain is present, the scaled excess energy density $L(e_z-e_{z,0})$ equals $4$ if the domain walls are smooth. The plateau value
in Fig.~\ref{fig:dE}(b) tends to a value close to $1.5$, which implies that about 40\% of the configurations resulting from these SA processes host stripe
domains (with the approximation of smooth domain walls). The large number of such samples is consistent with previous studies of sudden quenches
\cite{spirin01a}, where random spin configurations ($T=\infty$) were evolved by MC updates at $T=0$. In this case, smooth horizontal or vertical domain
walls are frozen in and the fraction of such samples, which can be computed exactly \cite{barros09,olejarz12}, is about $0.34$. It was also found that
diagonal domain walls, such as those illustrated in Fig.~\ref{fig:config}(c), vanish on a time scale $L^a$ with $a \approx 3.5$ at fixed $T<T_c$
\cite{olejarz12}, surprisingly being different from the $L^3$ scale of elimination horizontal or vertical domains \cite{lipowski99} at low $T>0$.
The almost perfect data collapses versus $vL^3$ in Figs.~\ref{fig:m2all}(c) and \ref{fig:dE}(b) suggests that the value of the exponent is exactly
$3$. In Sec.~\ref{sec:wind} we will demonstrate explicitly that this $L^3$ scaling indeed involves only $W=(1,0)/(0,1)$ domains.

\begin{figure}[t]
\includegraphics[width=7.5cm,clip]{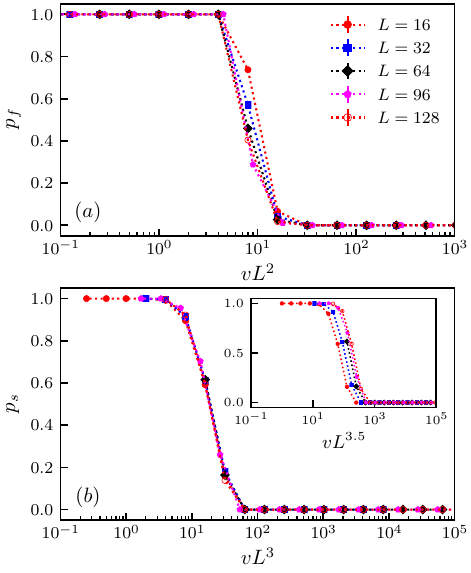}
\vskip-1mm
\caption{Probability of reaching one of the perfectly ferromagnetic ground states for the same fast (a) and slow (b) annealing instances as in
Fig.~\ref{fig:m2}, graphed against the corresponding scaled velocities $vL^2$ and $vL^3$.}
\label{fig:prob}
\end{figure}

The $L^2$ and $L^3$ scaling behaviors can be further refined by dividing the final spin configurations into two groups according to their $m^2$ values.
A value close to $1$ typically means that no large-scale defect is present, and then the $L^2$ scaling should be more prominent. Conversely, by analyzing
the group with small $m^2$ the $L^3$ scaling should be better emphasized. Here we define the groups to contain 20\% of the total number of configurations; those
with the largest and smallest $m^2$ values, which we refer to as the ``fast'' and ``slow'' groups, respectively. As shown in Fig.~\ref{fig:m2}(a), the
fast group does not show any deviations from the $L^2$ time scaling, apart from some scaling corrections for the smaller systems, thus indicating that
there are no instances of stripe domains in this group. In Fig.~\ref{fig:m2}(b), there is still $L^2$ scaling when $m^2$ is below the value of the
plateau that builds up with increasing $L$, which is natural since the configurations with stripe domains also contain confined domains at high velocity
and coarsening is also the mechanism by which a fractal domain in the neighborhood of $T_c$ evolves to a stripe  with smoother interfaces in the
ordered phase. The ultimate saturation of the order parameter to $m^2=1$, which takes place on the time scale $L^3$, is still more rapid than in
Fig.~\ref{fig:m2all}(c) because of less surviving confined domains in the slow group.

Another metric for identifying the relevant time scales is to use the probability of reaching one of the perfect ground states, i.e., the fraction
of samples with $m^2=1$ after annealing. Fig.~\ref{fig:prob} shows this probability in the fast and slow groups, again versus $vL^2$ and $vL^3$,
correspondingly. The good data collapses further validate the $L^2$ and $L^3$ scales. The inset of Fig.~\ref{fig:prob}(b) shows a contrasting plot with
the scaling $vL^{3.5}$ expected with the relaxation time observed in sudden quenches \cite{spirin01a} (our improved exponent $a\approx 3.42$ for diagonal
stripe elimination, determined in Sec.~\ref{sec:wind}, does not alter the plot significantly). In principle, there may be a small correction to the
fully saturated magnetization that would be visible in high-precision results for $1-m^2$. However, we do not have sufficiently many $v$ values and
good enough statistics for such an analysis.

It has been argued that the time scale of a straight stripe domain to vanish is $\propto L^3{\rm e}^{4J/T}$ in the limit of fixed low temperature
\cite{spirin01a}: Starting from perfectly straight domain walls, an indentation in one wall, corresponding to a probability
$\propto {\rm e}^{-4J/T}$ of a single-spin flip, provides a seed by which one layer of the defect domain can be peeled off by a random walk without
additional energy cost, thus requiring of the order $L^2$ steps to completely remove one layer. Removing of the order $L$ layers then leads to
the $\propto L^3$ time scale. While this appears a plausible scenario at extremely low temperatures, so that no additional seed defects form within
the time $\propto L^3$ taken to eliminate a domain, in reality multiple defects will be present in fluctuating domain walls at higher temperatures.
With SA spanning high to low temperatures, the above arguments do not appear to be applicable.

To study the way a stripe domain vanishes, we can identify events during the SA process where the winding number changes from non-zero to zero, which
marks the elimination of the last system spanning defect. The topological winding number of a configuration can be computed by assigning positive and
negative $x$ or $y$ unit currents to all lattice links associated with domain walls (i.e., those with different spin orientation on either side of the
link). Walking along a domain wall from an arbitrary initially selected location, the currents are accumulated until the same location is reached again,
repeating such loop-building processes for all domain boundaries in the system. A system without topological defects has no loops with net current,
while the current of a topological domain wall is a multiple of the system length $L$. We normalize such that the winding number is $W=(w_x,w_y)$
with positive components $w_x,w_y \in \{0,1,2,\ldots\}$. Note that there are restrictions on the winding numbers, e.g., $W=(2,0)$ is not possible.
The probability  $P(w_x,w_y)$ decreases rapidly with increasing winding, as will be discussed quantitatively in Sec.~\ref{sec:wind}.

\begin{figure}[t]
\center{\includegraphics[width=7.5cm, clip]{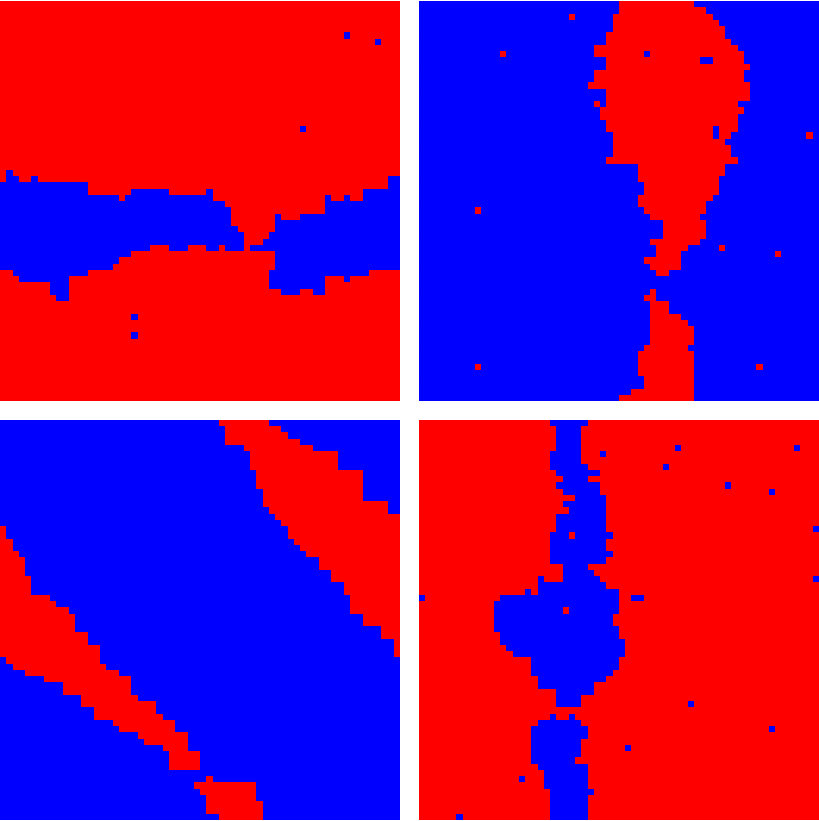}}
\caption{Snapshots of $L=64$ spin configurations generated by SA shortly after the winding number changed from non-zero to
zero. The minority domains have been cut by the majority species ``tunneling'' through via interface fluctuations. Events taking place
below $T=1.5$ were singled out.} 
\label{fig:cut}
\end{figure}

Applying the above method to identify configurations when their winding number change from non-zero to zero (in practice checking the winding after each
full MC sweep), we further apply a temperature $T^*$ at which we demand non-zero winding and only output configurations after zero winding is found during
the subsequent part of the SA process down to $T=0$. Fig.~\ref{fig:cut} shows examples of such recently broken stripe domains generated for a system with
$L=64$ and $T^*=1.5$. The winding number changed to zero only slightly below $T^*$ in all cases shown. Here it is clear that the still rather thick
minority domains have been pinched off by the two interfaces colliding, which is also the mechanism suggested in Ref.~\cite{lipowski99} at fixed $T<T_c$.
While the resulting severed domain may also have been thinning overall before it breaks, there are no long straight segments where the mechanism of diffusion
of a single seed defect over large distances can take place. Thus, the $\propto L^3$ time scale that we observe should have a different origin than the
low-$T$ mechanism proposed in \cite{spirin01b}. A related problem of an interface model driven by deposition and evaporation of particles \cite{plischke87},
mapped to a kinetic Ising model, can account for the $L^3$ scaling, but the exponent is not universal.

We do not observe any signs of a time scale $\propto L^a$ with $a \approx 3.5$ in $m^2$ or $e_z$, which suggests that the diagonal stripes in SA vanish
faster than those in instantaneous quenches to $T=0$ \cite{spirin01a}. We will show in Sec.~\ref{sec:wind} that the presence of diagonal domains in SA is
actually controlled by the KZ mechanism, the time scale $L^{3.17}$ of which is faster than the corresponding $\propto L^{a}$ time to eliminate a diagonal
domain at fixed $T < T_c$. We will also obtain a more precise estimate of the exponent in fixed-$T$ simulations; $a= 3.42 \pm 0.02$. If the time
scale in SA indeed is $L^{3.17}$, we would still expect some impact on $m^2$ or $e_z$, specifically a third regime of scaling collapse if the lowest-$v$
data are graphed versus $vL^{3.17}$ instead of $vL^3$, i.e., the same data should then be described by three different scaling functions (of $vL^2$,
$vL^3$, and $vL^{3.17}$). However, diagonal domain are relatively rare at $T=0$ and the presence of scaling corrections and statistical errors make
it difficult to separate the two cases with rather similar exponents $3$ and $3.17$. The winding-number method used in Sec.~\ref{sec:wind} completely
isolates the configurations with diagonal domains and enables us to study their decay time more precisely.

To further characterize the path of a system to full ordering via removal of large-scale domains, it is useful to study Fourier modes of the spin texture
at small non-zero wave-vector ${\bm q}$;
\begin{equation}\label{mq}
m_{\bm q} =\frac{1}{N}\sum_{x,y}\sigma_{xy} e^{-i(xq_x+yq_y)},
\end{equation}
where we now have labeled the Ising spins by their lattice coordinates, $x,y \in [0,L-1]$. The expectation value of $m_{\bm q}$ vanishes by symmetry but
its absolute value squared is useful as a generalization of $m^2$. We here take $k$ to mean the longest wavelength along either the $x$ or $y$ axis,
$k_x=(2\pi/L,0)$ and $k_y=(0,2\pi/L)$, and take the sum of the two squares,
\begin{equation}\label{mk}
m^{2}_k = \langle m^*_{k_x}m_{k_x}\rangle + \langle m^*_{k_y}m_{k_y}\rangle,
\end{equation}
which can indiscriminately detect system-spanning domains forming parallel to either the $x$ or the $y$ axis. For such a stripe of relative width $x/L$
(which we here treat as a continuous variable even for finite $L$) and completely smooth domain walls, the uniform magnetization is
\begin{equation}
m=\pm \left (1- \frac{2x}{L}\right ),
\end{equation}
while the squared long-wavelength order parameter is
\begin{equation}
m_{k}^2=\frac{2[1-\cos(\frac{2\pi x}{L})]}{\sin^2(\frac{\pi}{L})L^2}.
\end{equation}
By expressing $x/L$ as a function of $m^2$ and then rewriting $m_{k}^2$ in terms of $m^2$, we obtain the following form of $m_{k}^2$ as a function of $m^2$:
\begin{equation}\label{mkvsm}
m^{2}_{k}=\frac{2[1-\cos(\pi(\sqrt{m^2}+1))]}{\sin^{2}(\frac{\pi}{L})L^2}.
\end{equation}
While typical domain walls are not smooth, the limiting case is still useful. We have also calculated similar functional relationships between $m^2$
and $m^2_k$ for spin configurations with a single perfectly square or circular defect.

\begin{figure}[t]
\center{\includegraphics[width=7cm,clip]{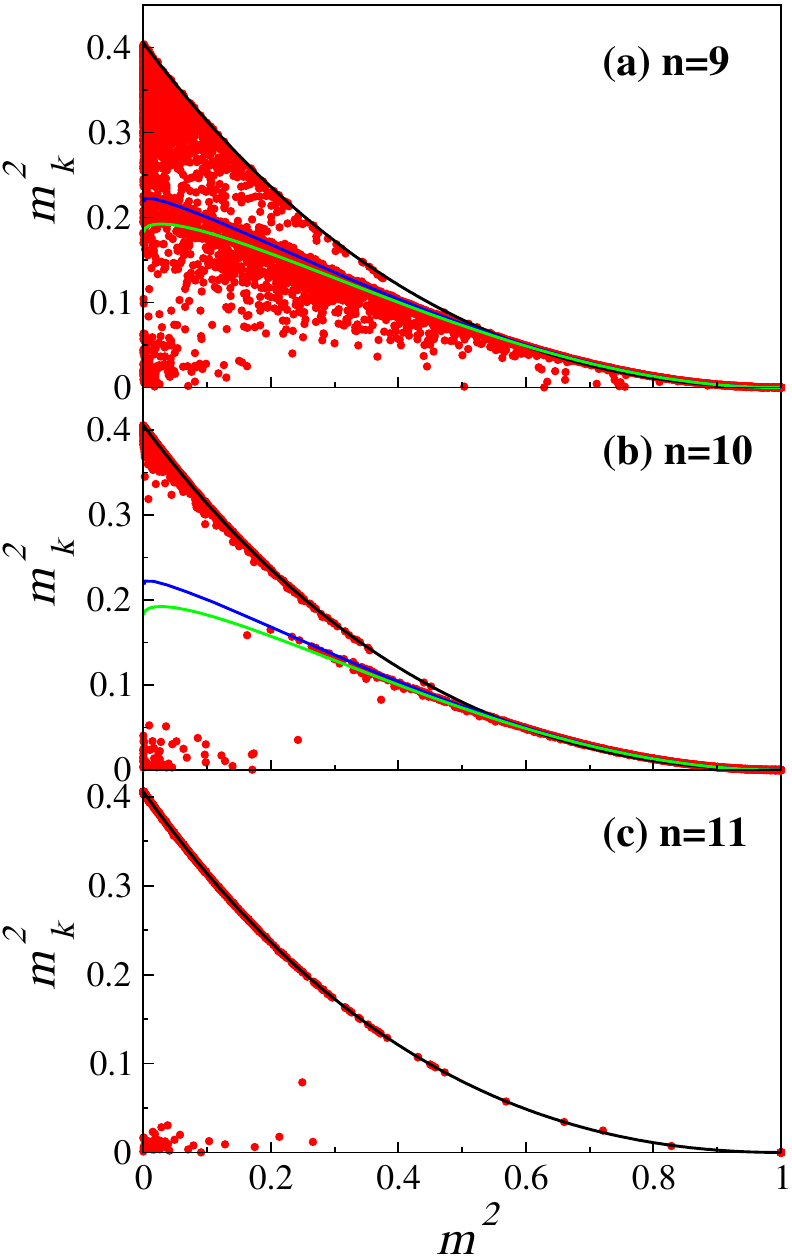}}
\caption{Scatter plots of the domain order parameter $m^{2}_{k}$, Eq.~(\ref{mk}), vs the conventional order parameter $m^2$ for thousands of
instances of an $L=64$ system annealed from $T=4$ to $T=0$ in $2^n$ steps with $n=9$ in (a), $n=10$ in (b), and $n=11$ in (c). The black curves
show the function Eq.~(\ref{mkvsm}) for systems with perfectly straight stripe domains, while the blue and green curves in (a) and (b) indicate,
respectively, such relationships for systems with single circular and square shaped domains. In (c), most of the instances ended at
$m^2=1$, $m^2_k=0$. The points with small $m^2$ and $m^2_k$, in the low-left corners, correspond predominantly to systems
with a remaining diagonal stripe defect.}
\label{fig:corr}
\end{figure}

Figure \ref{fig:corr} shows scatter plots of $m^{2}_{k}$ versus $m^2$ for $L=64$ systems annealed over $2^n$ MC steps with three different values
of $n$. The functional forms of $m^2_k$ versus $m^2$ for the straight system-spanning domains as well as the circular and square-shaped confined
domains are also shown. For long SA processes, most instances land on the point $(m^2,m^2_k)=(1,0)$ representing the two uniform ferromagnetic ground
states. As expected, Eq.~(\ref{mkvsm}) for smooth domains represents the upper bound for $m^2_k$ given $m^2$. The instances containing straight domain
walls concentrate closer to this bound as the annealing time is increased, especially in Fig.~\ref{fig:corr}(c), showing that the remaining domain
walls become smoother as the annealing time is increased. In Figs.~\ref{fig:corr}(a) and (b), many points also are accumulated close to the curves
representing square and circular domains, with the former providing an upper bound for all confined domains. However, in Fig.~\ref{fig:corr}(c) there are
no configurations left in the region of confined domains, while many instances of straight domains remain, reflecting the longer time required
for eliminating the latter in the instances where they are present. There are also points near the low-left corner of the diagrams, which correspond
mainly to configurations with diagonal stripe domains, for which both $m^2$ and $m^2_k$ vanish in the ideal case of smooth domain boundaries.
These domains could also in principle be analyzed as above by using the appropriate wave-number ${\bm q}$ in Eq.~(\ref{mq}).
Instead, in Sec.~\ref{sec:wind} we will analyze the probability distribution of the topological winding number. 

\subsection{System-spanning domains at fixed $T$}
\label{sec:wait}

Before further studies of the SA life time of system-spanning domains, we return to the question of these time scales at fixed $T <T_c$. To this
end, we initialize the system in a state of two domains, with $3/4$ the spins up and $1/4$ down and smooth domain walls, either horizontal
(the case first studied in Ref.~\cite{lipowski99}) or diagonal (studied at $T=0$ in Ref.~\cite{spirin01a}). These spin configurations have winding
number $W=(1,0)$ and $(1,1)$, respectively. Our aim is to confirm that these domains are really eliminated on different time scales, by studying
larger systems than previously. We perform MC simulations at fixed temperature significantly below $T_c$ and compute the the winding number after each
MC sweep. If $W=(0,0)$, which indicates that there is no longer a system-spanning defect domain in the system, the simulation stops and the number
$\tau_{\rm d}$ of MC sweeps performed is recorded. This procedure is repeated many times for a mean decay time $\langle\tau_d\rangle$. In addition,
the logarithm of $\tau_d$ is also averaged for an estimate ${\rm exp}\langle \ln(\tau_d)\rangle$ of the typical time. 

\begin{figure}[t]
\includegraphics[width=80mm]{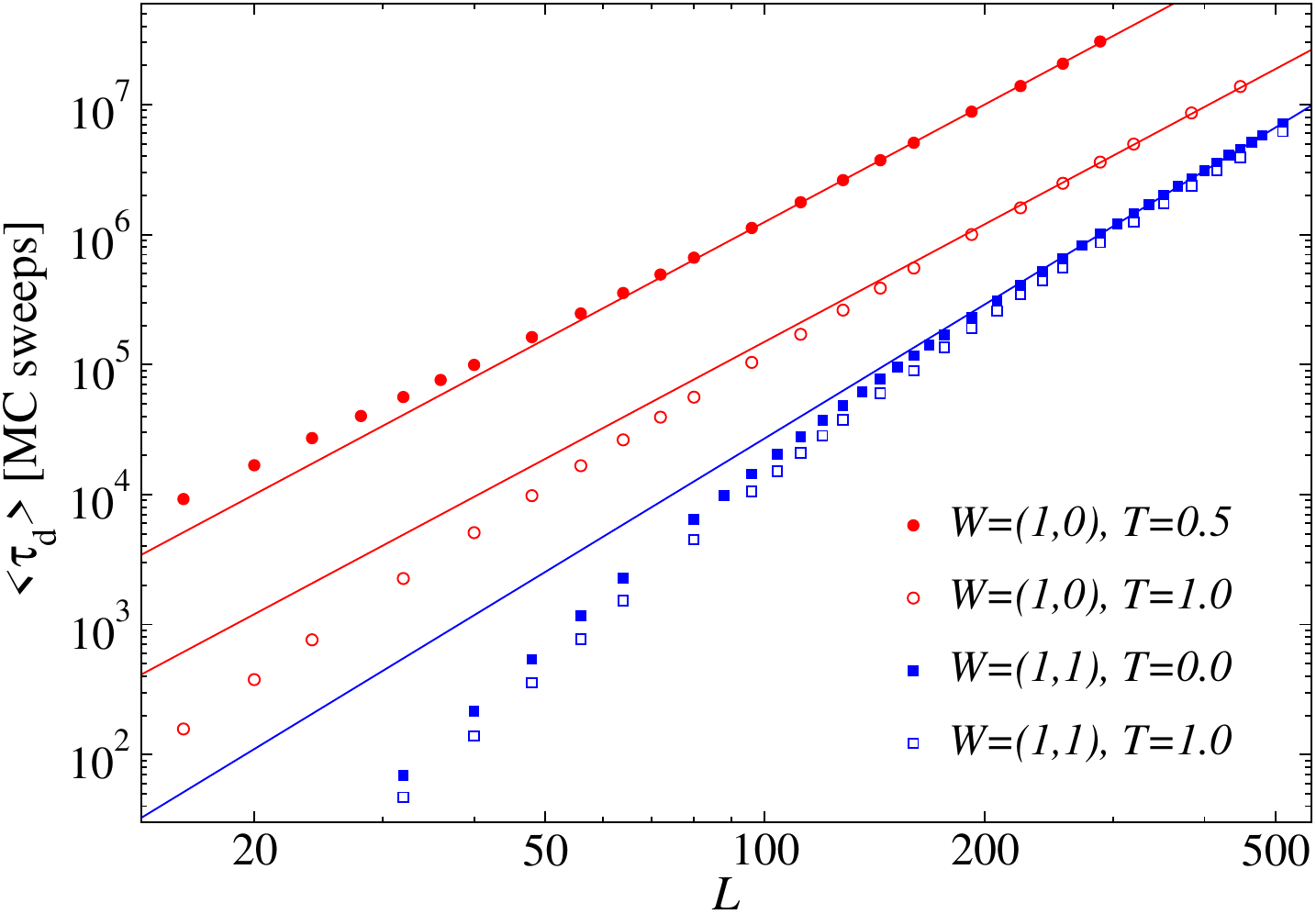}
\caption{Mean time required at fixed $T < T_c$ to remove a system-spanning minority domain (consisting of $1/4$ of the spins) imposed in the starting
configuration. Size dependent results are shown at two different temperatures each for horizontal, $W=(1,0)$, and diagonal, $W=(1,1)$, domains. The red
lines drawn through the $W=(1,0)$ large-size data correspond to $L^a$ power laws with $a=3$. The blue line is a fit to the $W=(1,1)$ data at $T=0$
for $L \ge 288$, with the resulting exponent $a\approx 3.42$. All error bars are much smaller than the plot symbols.}
\label{fig:healing_time}
\end{figure}

The mean decay time is graphed versus the system size in Fig.~\ref{fig:healing_time} for both types of domains at two different temperatures. In
the case of $W=(1,0)$ the expected form $\langle \tau_{\rm d}\rangle \propto L^a$ with $a=3$ is fully consistent with the large-$L$ data. A fit to the better
size-converged $T=0.5$ data set gives $a=2.987 \pm 0.013$ (the error indicating one standard deviation of the mean) when using data for $L \ge 128$
(for a statistically good fit). Fitting to the typical decay time (data not shown here) for the same range of system sizes gives a statistically equal
exponent, $a=3.008 \pm 0.014$, thus showing that the distribution of decay times does not have broad tails. The results for $T=1$ are also fully
consistent with $a=3$ but, because of the slower convergence to the asymptotic form, the statistical errors are larger.

Note that the prefactor of the $L^3$ form (the constant shift on the log scale used in Fig.~\ref{fig:healing_time}) for $W=(1,0)$ is much larger at the lower
temperature, which reflects the reduced domain-wall fluctuations with lowered $T$, with eventual freezing of domains with smooth walls at $T=0$. For $W=(1,1)$,
only a very weak dependence on the temperature is seen Fig.~\ref{fig:healing_time}, which is clearly related to the fact that the diagonal domains can decay
away also at $T=0$ (the lower temperature in Fig.~\ref{fig:healing_time}) and, apparently, non-zero temperature does  not significantly assist in the
large-scale interface fluctuations in this case. Fits to the $T=0$ data for $L \ge 288$ gives the exponent $a=3.420 \pm 0.015$ for the mean decay time
and $3.556 \pm 0.016$ for the typical time (for which data are not shown). The exponent for the mean time is roughly consistent with the less precise
result $a \approx 3.5$ in Ref.~\cite{spirin01a}, where the system sizes were smaller and the typical decay time was not addressed. We also tested whether
the data could potentially instead be described by $L^3$ with a multiplicative logarithm (also including an exponent on the log). However, the parameters
of this fit do not stabilize when the smaller sizes are gradually excluded. The $T=1$ results are also consistent with a pure $L^a$ form, with
the same exponents as the $T=0$ case; $a=3.432 \pm 0.024$ for the mean decay time and $a=3.564 \pm 0.024$ for the typical time, again from fits to
the $L \ge 288$ data. Thus, for the diagonal domains it appears that the mean and typical time scales are different. Both the mean and typical
exponents for diagonal walls are larger than the KZ exponent $z+1/\nu \approx 3.17$, which is very significant in the context of SA, as will be
discussed next.

\subsection{Winding number analysis}
\label{sec:wind}

\begin{figure}[t]
\includegraphics[width=84mm]{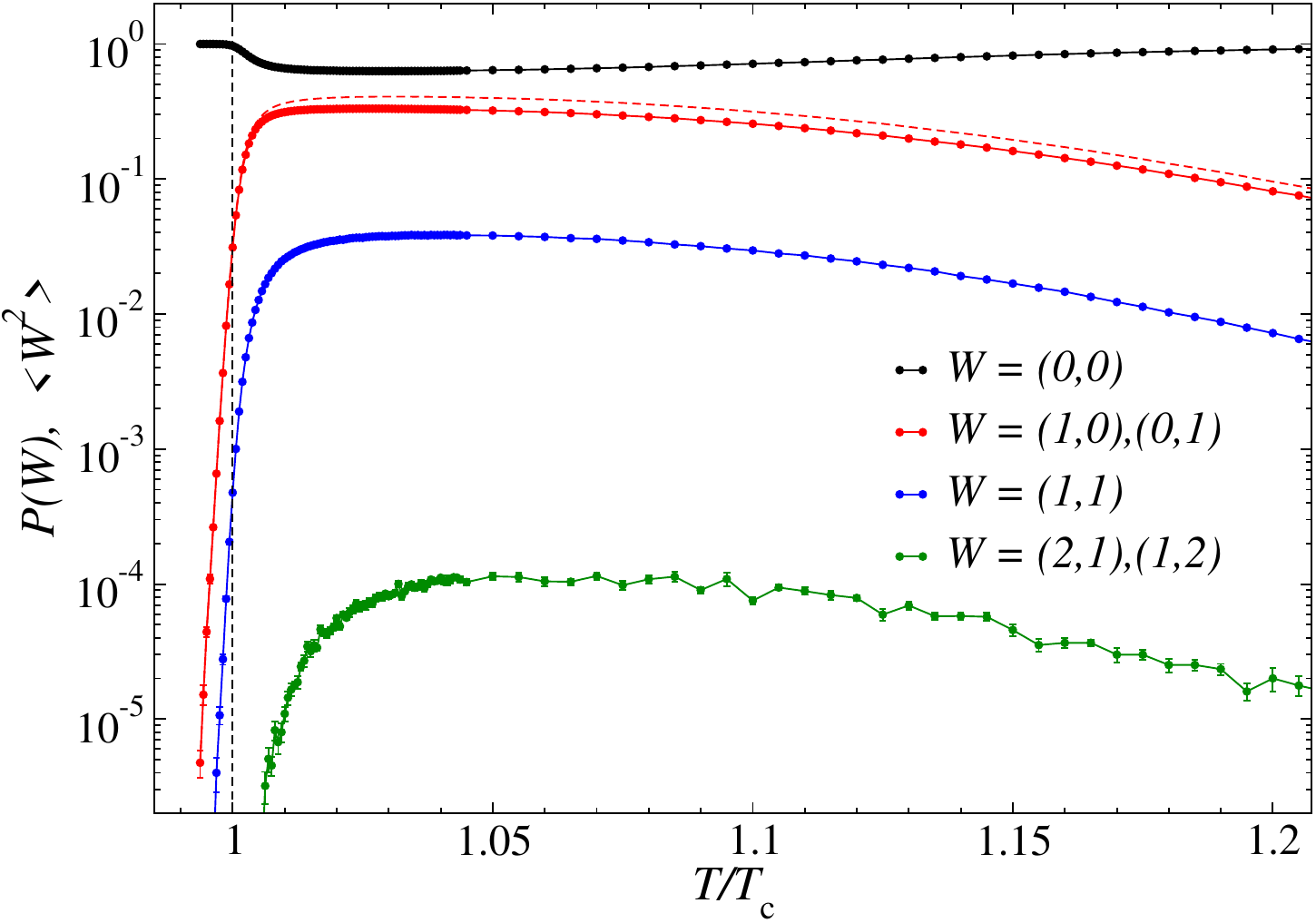}
\caption{Temperature dependent probabilities of the four dominant winding number sectors obtained by equilibrium SW simulations for lattice size
$L=512$. The $w_x \not= w_y$ results represent the sum of the two equivalent probabilities. The red dashed curve is the mean squared winding number.}
\label{fig:pw}
\end{figure}

The fact that the time scale $L^3$ for elimination of straight domain walls below $T_c$ is shorter than the KZ scale $L^{3.17}$ implies that their
survival probability in the ordered phase is not governed by KZ scaling. Conversely, KZ scaling can be expected to apply to the persistence of
diagonal domain walls, since their time scale for elimination at $T < T_c$ is longer, of order $L^{3.42}$. This is a very interesting prospect in the
context of the original KZ ideas \cite{kibble76,zurek85}, which concerned large-scale topological defects akin to the system-spanning domain walls in the
Ising model. In addition to confirming KZ scaling for the $W=(1,1)$ diagonal domains in $T\to 0$ SA, we will here also show additional evidence for the
faster scaling of the $W=(1,0)/(0,1)$ horizontal/vertical domains. We do this by explicitly studying probabilities of different winding numbers, first
in equilibrium MC simulations and then with SA. Strictly speaking, the probabilities presented below were computed on the basis of the largest
winding number of an individual domain wall found in the generated spin configurations, i.e., $P(W)$ is the probability of $W=(w_x^2+w_y^2)^{1/2}$ being
the largest winding number among any of the domain walls in a sampled configuration. The probability of more than one winding domain in the same
configuration is very low, a manifestation of which is that the probabilities in the four winding sectors reported below almost completely
exhaust the mean of the squared winding number computed with all domain walls.

For the equilibrium simulations the  Swendsen-Wang (SW) cluster algorithm \cite{swendsen87} was used for system sizes up to $L=512$. The temperature dependence
of the probabilities of the sampled configurations to fall in one of the four most important winding number sectors is shown in Fig.~\ref{fig:pw} for the
largest system studied, along with the mean squared winding number. All $W \not= (0,0)$ probabilities vanish for $T \to \infty, L\to \infty$ because of the
zero percolation probability in the paramagnetic phase. The winding probabilities also have to vanish for any $T$ below $T_c$ in the thermodynamic
limit because of the large $\propto L$ internal-energy cost of system-spanning domain walls. There is significant size dependence overall, but the peak
values in Fig.~\ref{fig:pw} are close to converged, and also close to the exact probabilities \cite{olejarz12} (slightly larger for non-zero winding)
in quenches of random states to $T=0$ (and there is no reason for these probabilities to be exactly the same). The non-zero $W$ probabilities indeed
drop very rapidly in the neighborhood of $T_c$.

\begin{figure}[t]
\includegraphics[width=82mm]{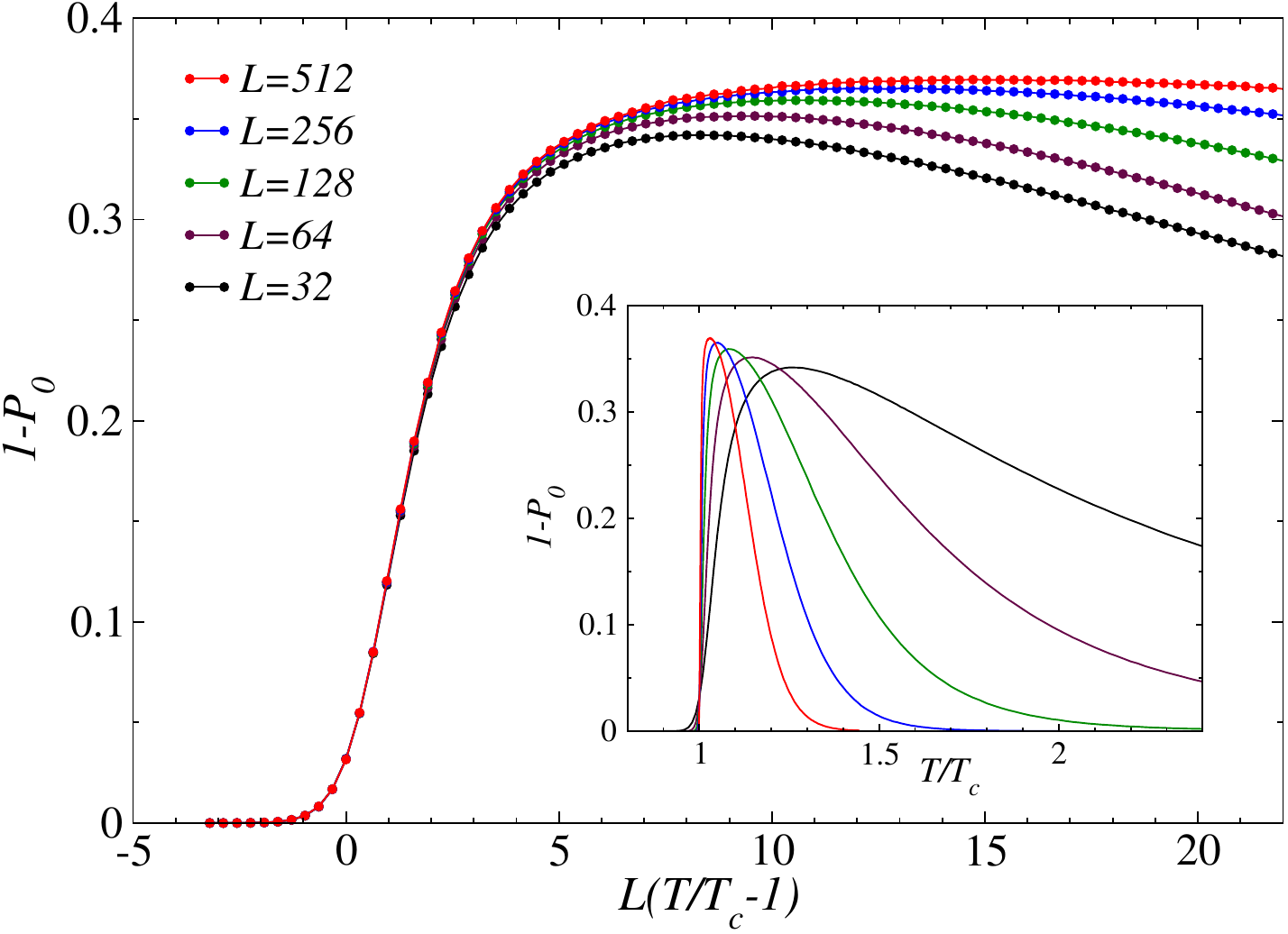}
\caption{Critical scaling with the 2D Ising exponent $\nu=1$ of the probability of non-zero winding number in equilibrium SW simulations. The inset shows
the raw data before scaling, with the same color coding as in the main graph. Error bars are too small to be visible.}
\label{fig:p0}
\end{figure}

All the distributions $P(W)$ exhibit critical scaling in the form expected for a dimensionless quantity. Fig.~\ref{fig:p0} shows $1-P(0,0)$, i.e., the
probability of a spin configuration having at least one winding domain, versus the distance $\delta=T/T_c-1$ to the critical point scaled by $L^{1/\nu}=L$.
The data collapse well onto a common scaling function $p(\delta L)$ close to $T_c$, and a constant behavior emerges for large $\delta L$ with increasing
system size. The suppression of winding in the neighborhood of $T_c$ is again very prominent. We note that the extended domain walls existing in the
2D Ising model at $T_c$ are fractal and can be described by conformal field theory \cite{aasen16,hauru16}.

\begin{figure}[t]
\includegraphics[width=84mm]{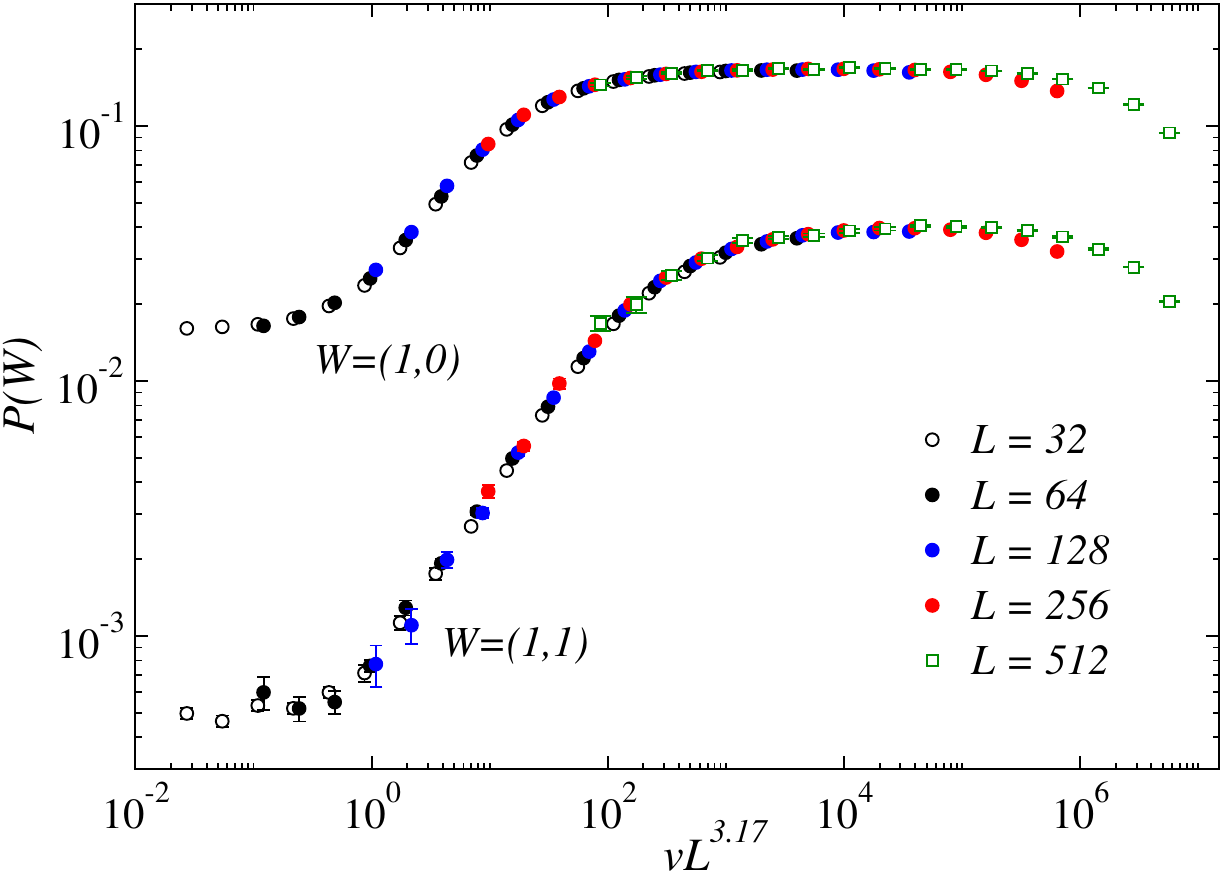}
\caption{Scaling of the probability of winding numbers $(1,0)$ and $(1,1)$ in SA runs ending at $T_c$ for five different system sizes. The
velocity has been scaled with the system size according to the expected KZ exponent $z+1/\nu \approx 3.17$.}
\label{fig:wtc}
\end{figure}

Next, we turn to the out-of-equilibrium simulations, again performed with standard Glauber dynamics. In this case the SA starting temperature was
$2T_c$. KZ scaling of the winding probabilities at $T_c$ is confirmed in Fig.~\ref{fig:wtc} with the examples of $W=(1,0)$ and $(1,1)$. Here $P(1,0)$ refers
to only $W=(1,0)$, though the equivalent $(0,1)$ probability was used for averaging. Note that the maximum values of the velocity-scaled probabilities in
Fig.~\ref{fig:wtc} are very close to the peak values slightly above $T_c$ in the corresponding equilibrium results for $L=512$ in Fig.~\ref{fig:pw}
[adjusted by the factor 2 in the case of $W=(1,0)$]. This behavior reflects the similarity of the near-critical state at mildly elevated $T>T_c$
($T-T_c \sim L^{-1/\nu}$) in equilibrium and the slightly out-of-equilibrium state in SA when stopping at $T_c$---both temperature and velocity effects
enhance the presence of system-spanning winding domains.

\begin{figure}[t]
\includegraphics[width=75mm]{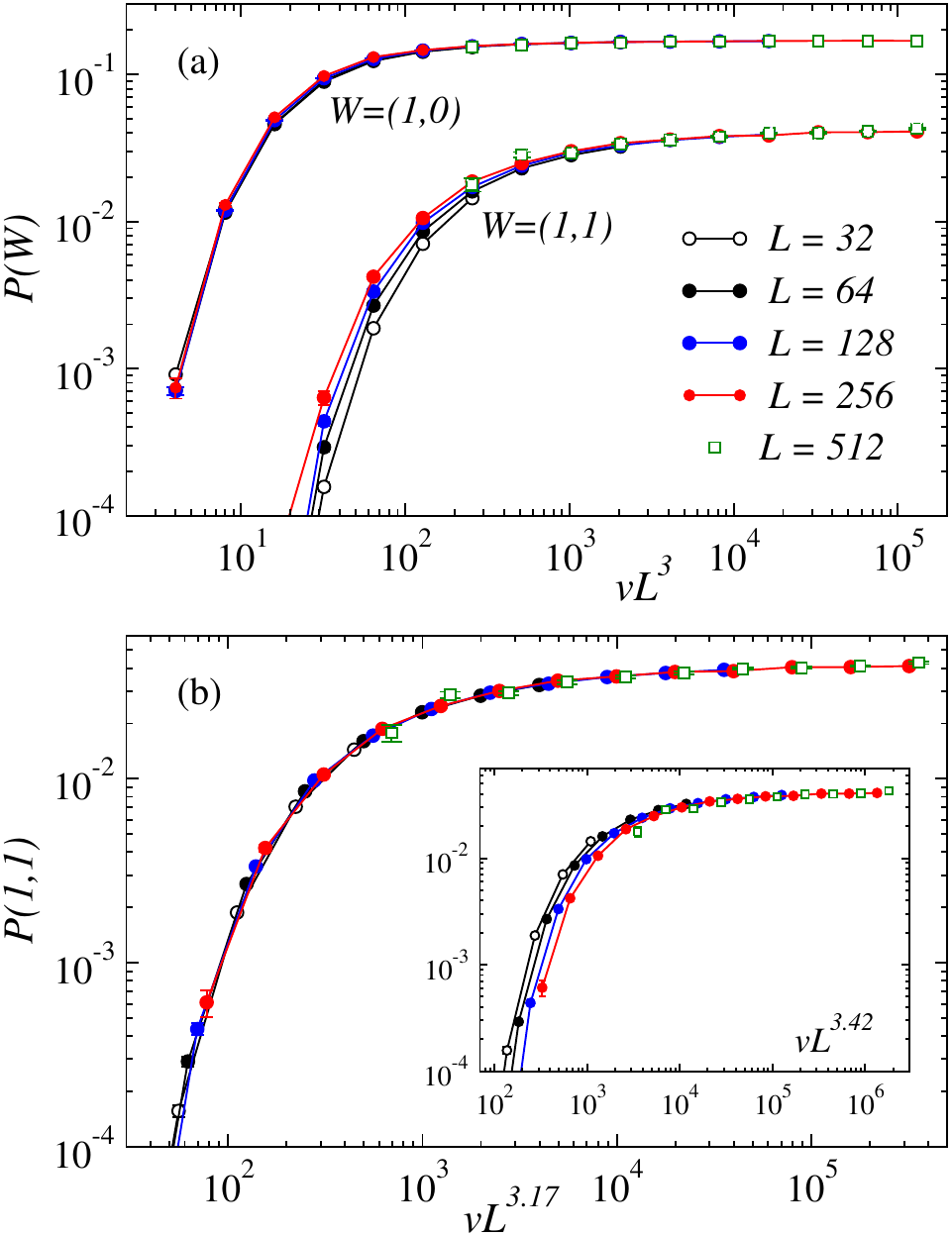}
\caption{Scaling with $v$ and $L$ of the probability of winding numbers $(1,0)$ and $(1,1)$ after SA to $T=0$. In (a), both data sets are graphed vs $vL^3$, which
leads to good data collapse for the five system sizes considered only for $W=(1,0)$. Note that many of the $L=512$ data points over those for the smaller systems.
In (b) the $W=(1,1)$ data are instead collapsed vs $vL^{3.17}$, corresponding to KZ scaling. The inset of (b) demonstrates the absence of data collapse when
$v$ is scaled by $L^{3.42}$; the fixed-$T$ scale of diagonal domain elimination.}
\label{fig:wt0}
\end{figure}

\begin{figure}[t]
\includegraphics[width=75mm]{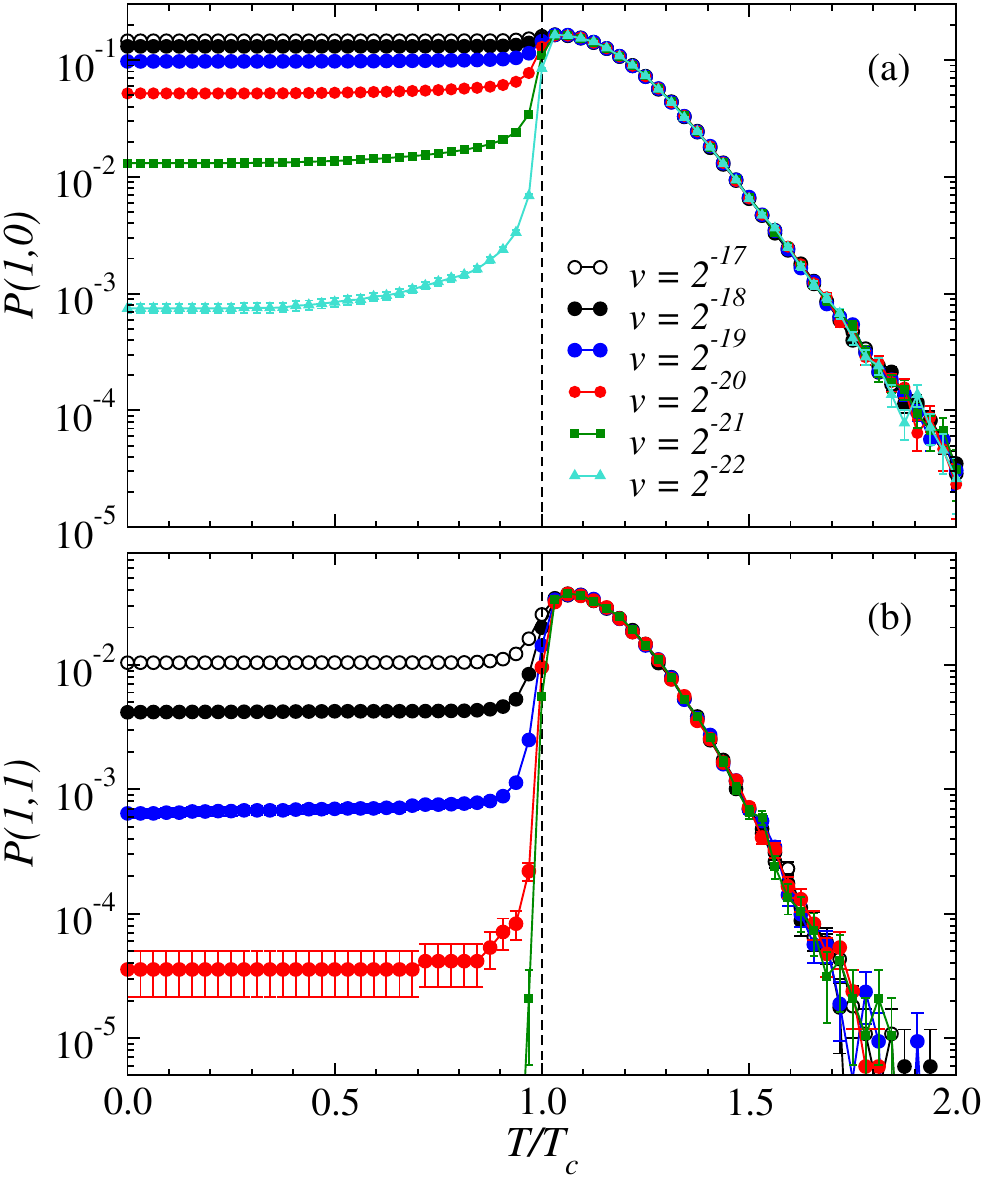}
\caption{Temperature dependent probability of winding number $(1,0)$ (a) and $(1,1)$ (b) during SA from $T=2T_c$ to $T=0$. The results are
for fixed system size $L=256$ and different velocities $v$, with $v^{-1}$ being the total annealing time.}
\label{fig:pvst}
\end{figure}

For SA continuing all the way to $T=0$, Fig.~\ref{fig:wt0}(a) reveals the $L^3$ scaling of the $W=(1,0)$ probability, as should be expected based on
the results in Sec.~\ref{sec:sascale}. However, the $W=(1,1)$ data do not collapse well versus $vL^3$. Instead, scaling the velocity with the KZ time
$L^{z+1/\nu}$ produces excellent data collapse, as shown in  Fig.~\ref{fig:wt0}(b). This difference between the two winding sectors is consistent with the
time scale $L^{3.42}$ found in Sec.~\ref{sec:wind} for the elimination of the diagonal domain walls at fixed $T < T_c$. The KZ scale is shorter
and, therefore, the $W=(1,1)$ domains will primarily decay very soon after $T_c$ has been crossed (and even slightly before, judging from the
peak locations in Fig.~\ref{fig:pw}), with a slower decay rate during the part of the SA process from $T_c$ down to $T=0$ so that KZ scaling remains
valid even at $T=0$. Scaling with $L^{3.42}$ does not produce good data collapse (except in the almost flat portion that is not sensitive to small
variations in the exponent), as shown in the inset of Fig.~\ref{fig:wt0}(b).

The fast decay of the winding probabilities immediately below $T_c$ is confirmed in Fig.~\ref{fig:pvst}, which shows $L=256$ results accumulated after
each segment of $1/64$ of the total number of annealing steps. After the initial drop, the probabilities only decay slowly as $T$ is further lowered.
The slow evolution in the case of $W=(1,1)$ is perhaps surprising in light of the mean times to eliminate diagonal stripe domains at fixed $T$ in
Fig.~\ref{fig:healing_time}, which for $L=256$ is not much longer than the almost flat portion during the SA process at $v=2^{-12}$ in
Fig.~\ref{fig:pvst}(b). However, it should be kept in mind that the fixed-$T$ runs were started with rather thin domains (containing $1/4$ of
the spins), while the domains remaining after the initial steep decay in SA are those that are the most difficult to eliminate.
In the case of $P(1,0)$, the tendency to further decay at low temperatures is more obvious, especially in the case of the lowest $v$ in
Fig.~\ref{fig:pvst}(a). Because of the overall much smaller scale of the $P(1,1)$ probabilities, statistically meaningful $T\to 0$ results were
limited to higher velocities in Fig.~\ref{fig:pvst}(b).

\subsection{Open boundary conditions}
\label{sec:open}

\begin{figure}[t]
\includegraphics[width=75mm]{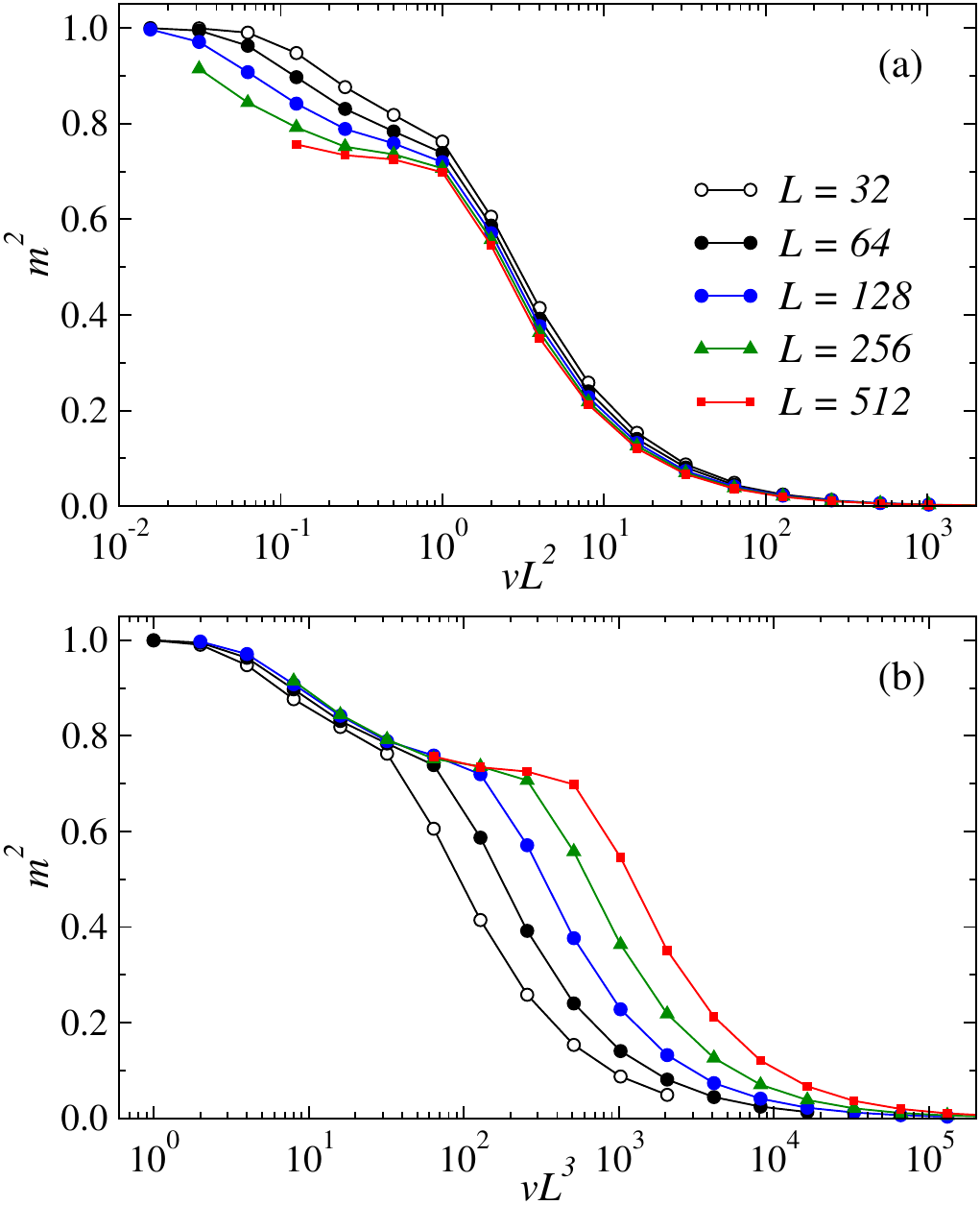}
\caption{Squared order parameter of systems with open boundary conditions at the end of $T \to 0$ SA runs for different velocities and system
sizes. The data are analyzed in the same way as those for the periodic systems in Figs.~\ref{fig:m2all}(b) and \ref{fig:m2all}(c), revealing the times
scales $L^2$ in (a) and $L^3$ in (b). Error bars ar at most of the size of the plot symbols (for $L=512$ at the lowest-$v$ points).}
\label{fig:open}
\end{figure}

The characterization of a system-spanning domain wall by a specific winding number is not possible in an open system. Moreover, in this case there
is only a single domain wall in a configuration with two domains. Nevertheless, the ends of such domain walls touch opposite edges of the system and
that also prohibits shrinkage of a domain in the ``long'' direction until an interior point of the domain wall reaches an edge, thus breaking
one of the domains up into two pieces. Alternatively, the end points of the domain wall can simply migrate toward each other on adjacent edges and
shrink one of the domains that way. It is not a priori clear whether the fluctuations of these long domain walls should be qualitatively different
from those in a periodic system, and what the time scale of elimination of system-spanning domains are.

Open boundary conditions are of interest in the context of QA experiments because periodic boundaries cannot be easily accessed with some of the
current platforms, such as Rydberg atom arrays \cite{manovitz25,zhang25}. With superconducting qubit arrays periodic boundary conditions can be implemented
by constructing logical qubits consisting of two physical qubits, as has been done with D-Wave devices \cite{king23}. A larger number of couplers in the most
recent D-Wave device allows for fully periodic boundaries of 2D square lattices with up to $12\times 12$ native qubits \cite{sathe25} (larger cylindrical
lattices, i.e., periodic in only one direction, have been also implemented \cite{king25}). It will still be useful to also investigate open boundaries. In SA,
we here focus on system-spanning domain walls and their time scales, and we expect the results to also be relevant in the context of QA (for which we will
only consider small periodic lattices in Sec.~\ref{sec:QA}).

We first confirm the presence of both the $L^2$ and $L^3$ time scales also in open systems by studying $m^2$ after SA to $T=0$, starting
at $T_{\rm ini}=2T_c$. Fig.~\ref{fig:open} shows scaled results analogous to those for periodic systems in Figs.~\ref{fig:m2all}(b) and
\ref{fig:m2all}(c). The data for open systems collapse in the same way, with only slightly larger corrections, and overall the shapes of the collapsed
data (the scaling functions) look very similar. The $L^3$ scale is again presumably related to horizontal or vertical domain walls.

\begin{figure}[t]
\includegraphics[width=80mm]{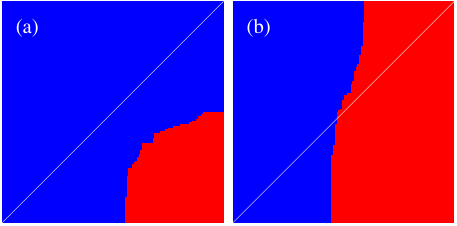}
\caption{Configurations from MC simulations of a system of size $L=128$ at $T=0$, obtained by starting from a configuration in which the spins below and
above the $x=y$ diagonal (the white lines) are all up (red) and all down (blue), respectively. The spins exactly on the diagonal were initially generated
at random. The fluctuations of the domain wall over time eventually leads to either (i) a fully ferromagnetic state, the fate of the case in (a),
or (ii)  a state with two domains separated by a stable (at $T=0$) horizontal or vertical smooth domain wall to which the (b) configuration will evolve.}
\label{fig:opendia}
\end{figure}

To specifically investigate the fate of diagonal domain walls in open systems,
we carry out calculations at fixed $T<T_c$ with a domain wall imposed initially, in analogy
with Sec.~\ref{sec:wait} for the periodic systems. Since a diagonal domain wall can be eliminated by local MC updates also at $T=0$ (and, moreover,
Fig.~\ref{fig:healing_time} shows only weak $T$ dependence for diagonal domains in periodic systems), we focus on that limiting case. A diagonal
domain wall is created by randomly generating up and down spins on the $x=y$ line on open $L\times L$ lattices and setting all the spins below (above)
this line to up (down). A spin can only be flipped if the energy is lowered or stays the same, and we carry out $N$ such attempts at random locations
for each time unit. This $T=0$ MC simulation could in principle be made much more efficient by only considering spins adjacent to the single
fluctuating domain wall, which we did not implement because very large systems are not
needed for our conclusions here.

The single domain wall, with end points that are initially on opposite corners of the square lattice, can evolve in two qualitatively different ways
that are illustrated with representative spin configurations in Fig~\ref{fig:opendia}: (i) The end points move to adjacent edges and continue to gradually
move toward each other while the length of the wall is reduced over time, thus creating a shrinking domain in one corner of the system
and eventually leading to the fully ferromagnetic state. Such a case of a shrinking corner domain is shown in Fig~\ref{fig:opendia}(a). (ii) The
end points move to opposite edges of the lattice and the domain wall becomes increasingly horizontal or vertical with time, which also shortens it
and reduces its energy from that of the initially diagonal wall. Eventually a completely straight wall is formed that cannot evolve further. A configuration
where such a stable domain wall ultimately forms is shown in Fig.~\ref{fig:opendia}(b). Whichever of the cases (i) and (ii) is realized depends on the very
early stages of the MC process. We have confirmed to high statistical precision that the probability is $1/2$ for both, as would be expected based on the
equal probabilities of initial fluctuations of the end points of the wall to either adjacent of opposite edges of the lattice.

\begin{figure}[t]
\includegraphics[width=80mm]{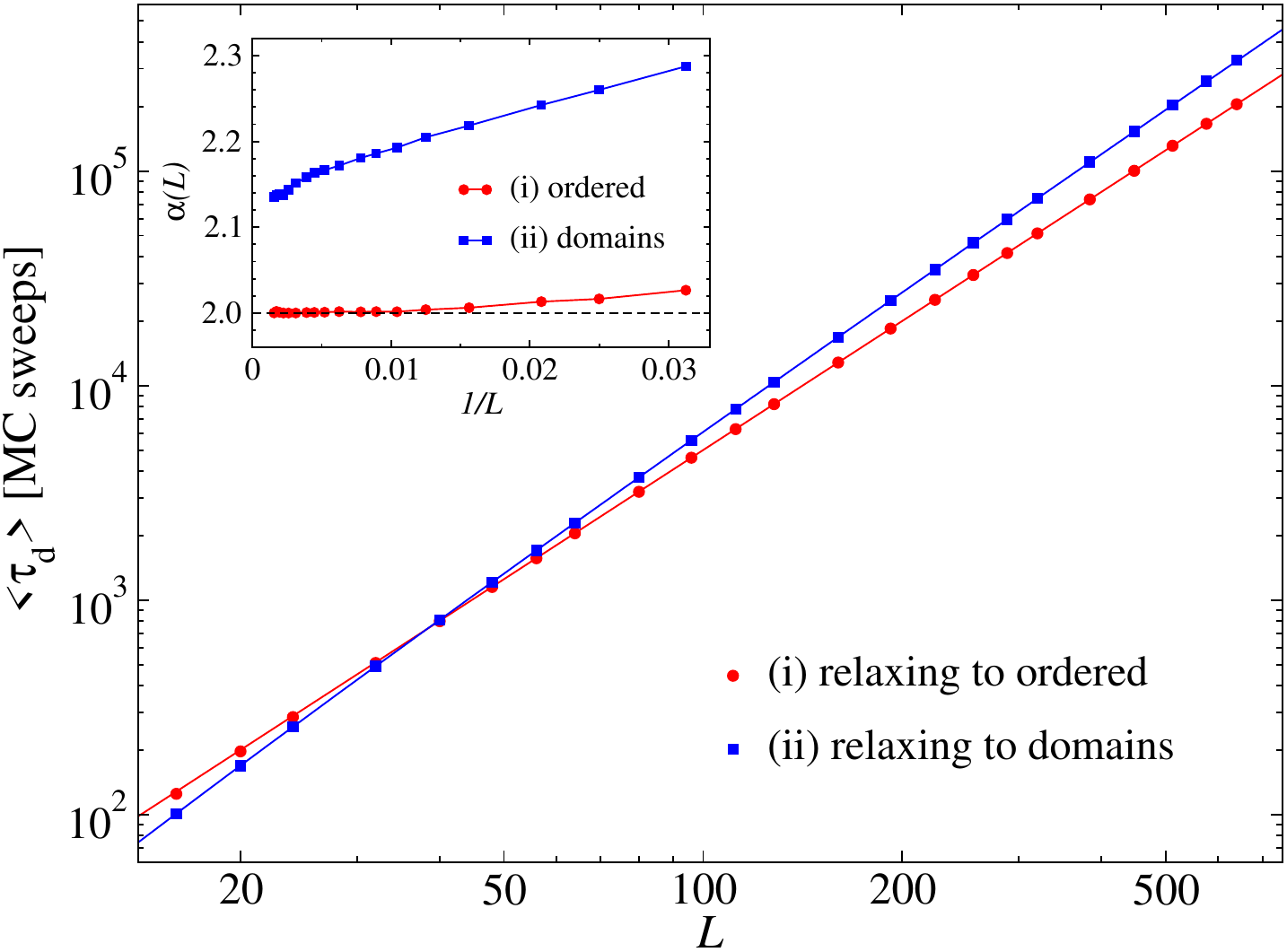}
\caption{Size dependence of the mean time at $T=0$ for systems with an initially imposed diagonal domain wall to evolve into (i) a fully ferromagnetic state
(red circles) or a (ii) stable state with a straight domain wall (blue squares), as illustrated in Fig.~\ref{fig:opendia}. The red fitted line has the form
$\langle \tau_{\rm d}\rangle \propto L^2$, while the blue curve is from a fit including also a multiplicative log correction;
$\langle \tau_{\rm d}\rangle \propto L^2\ln^a(L/b)$, with $a=0.72$ and $b=1.7$. Error bars are much smaller than the plot symbols. The inset shows
size dependent exponents $\alpha(L)$ obtained from data for pairs of system sizes $(L,2L)$ under the assumption of pure power laws
$\propto L^{\alpha}$. The results for case (ii) are consistent with a log correction to the exponent $\alpha(L)$ while in case (i) the
corrections are small and conventional.}
\label{fig:otime}
\end{figure}

Despite the system-spanning initial domain wall, the mean time to evolve to the fully ferromagnetic state in case (i) scales as $L^2$, i.e., the conventional
coarsening time, as shown with the red symbols and a fit in Fig.~\ref{fig:otime}. However, in case (ii) the time to reach a stable straight domain wall
is slightly longer and can empirically be described by $L^2$ with a multiplicative log correction, as shown with the blue symbols and the fitted
function in Fig.~\ref{fig:otime}. A plain log correction works for all but the smallest few system sizes, but with an exponent $\approx 0.7$ on the logarithm
also the smallest systems can be accounted for. The inset of Fig.~\ref{fig:otime} shows running exponents for both cases (i) and (ii)
obtained from data for system sizes $L$ and $2L$ by assuming the form $\langle \tau_d\rangle \propto L^\alpha$. Here a rapid convergence to $\alpha=2$ is
seen for (i), while for (ii) the values decrease slowly with increasing $L$ in a way consistent with a log correction. Thus, we conclude that, even in
the setting of open boundaries, the anomalous fluctuations of the diagonal domain wall survive, now leading to a correction to conventional coarsening,
even though in this case the state can only be classified as having a diagonal domain wall at the initial stages of the simulation. We note again that the
time scaling of diagonal stripe domains in periodic systems, Fig.~\ref{fig:healing_time}, is not well described by $L^3$ with a log correction but really
seems to demand a different power law; $\langle \tau_d\rangle \propto L^{3.42}$. 

\begin{figure}[t]
\includegraphics[width=80mm]{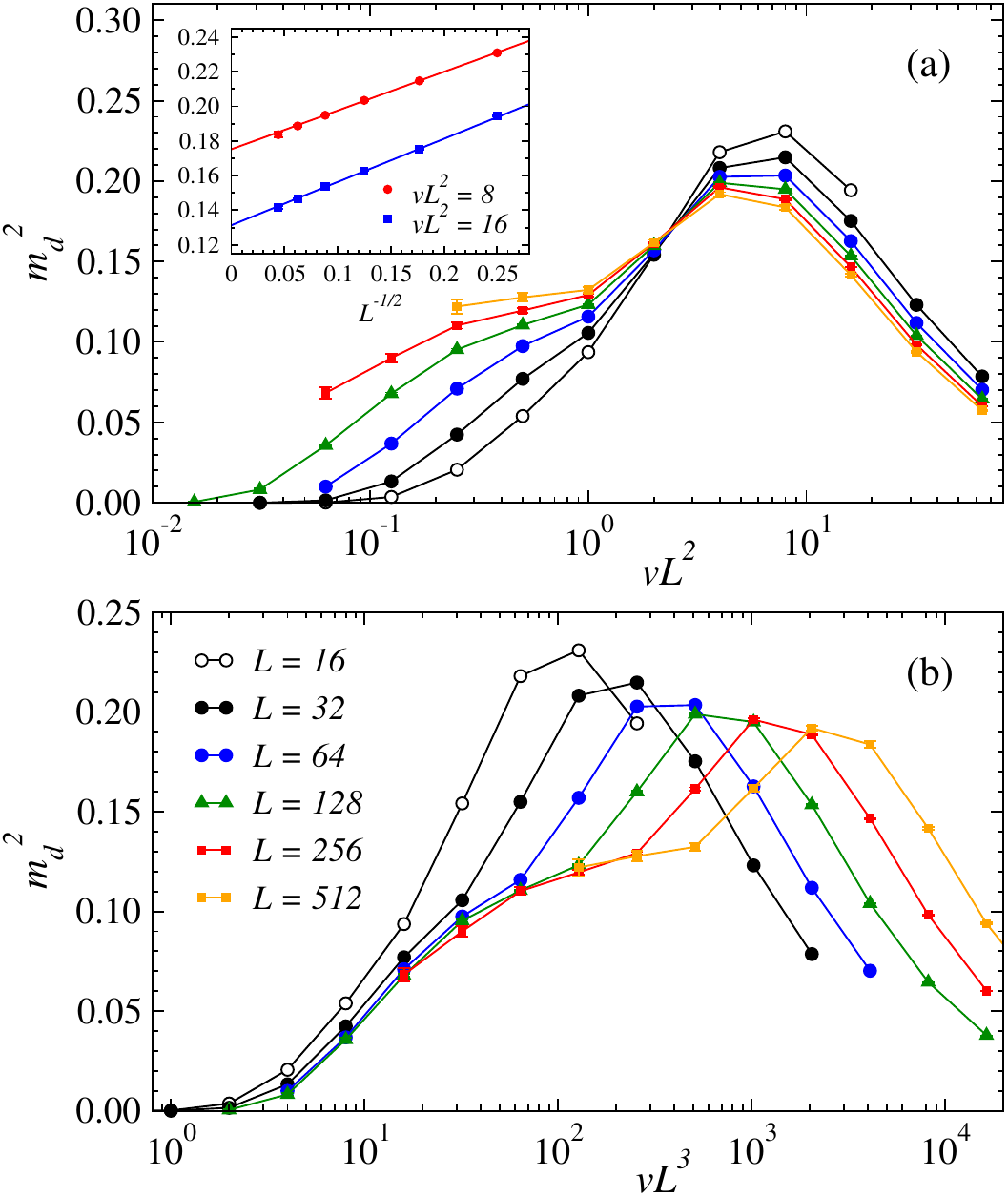}
\caption{The domain-wall order parameter defined in Eq.~(\ref{mddef}) computed in SA for different system sizes and velocities. The size legends in
(b) also apply to (a). The data are graphed with the velocity rescaled by the conventional coarsening time scale $L^2$ in (a) and by the scale $L^3$
corresponding to domain-wall elimination interface fluctuations in (b). In the inset, the data for $x=vL^2=8$ and $16$ are graphed versus the square-root
of the inverse system sizes along with fits to the form $a+bL^{-1/2}$ with free parameters $a$ and $b$.}
\label{fig:odef}
\end{figure}

While the resulting straight domain wall in case (ii) is stable at $T=0$, at $T>0$ the interface can again fluctuate, and once a point reaches an edge
of the system one of the domains is severed and can decay away by coarsening as in case (i). The time scale of the domain wall colliding with an
edge should be the same as that of of two walls colliding in the periodic systems, thus explaining the $L^3$ scaling in Fig.~\ref{fig:open}(b).
We can also explicitly confirm the origin of the $L^3$ scale in SA of open systems by defining a domain-wall order parameter similar to $m^2_k$ in
Eq.~(\ref{mk}). Here we simply transform the spins by functions $r_x(x,y)$ and $r_y(x,y)$ of unit magnitude that change signs when crossing the vertical
and horizontal central axis of the lattice, respectively, i.e., with $x,y \in \{0,\ldots,L-1\}$ we have $r_x(x<L/2,y)= 1$ and $r_x(x\ge L/2,y)= -1$
and similarly for $y$ in $r_y(x,y)$. The order parameter detecting a vertical domain wall is then defined as
\begin{equation}\label{mdxdef}
m_{dx} = \frac{1}{N}\sum_{x,y} r_x(x,y)\sigma_{x,y}, 
\end{equation}
along with an analogous definition of $m_{dy}$ to detect a horizontal domain wall. We compute the corresponding mean squared order parameter
\begin{equation}\label{mddef}
m_d^2 = \langle m_{dx}^2 \rangle  + \langle m_{dy}^2 \rangle. 
\end{equation}  

Figs.~\ref{fig:odef}(a) and \ref{fig:odef}(b) show results versus the velocity scaled by $L^2$ and $L^3$, respectively, averaged over the final $T=0$
configurations after SA starting at $T=4$.
Data collapse when plotting versus $x=vL^2$ is not completely apparent in Fig.~\ref{fig:odef}(a), though the peak value seems to converge with increasing
$L$ and the data on the right-hand size of the peak also seem to approach a common scaling function. Convergence is demonstrated in the inset of
Fig.~\ref{fig:odef}(a) with data at $x=8$ and $x=16$, where fits with square-root corrections to non-zero $L\to \infty$ values match the data very well.
The domains at $T_c$ are fractal in equilibrium and at low velocity, and data collapse when using the time scale $L^2$ should correspond to coarsening
dynamics initially bringing some of the SA instances toward a state with a non-fractal domain wall (and the instances with only confined defects will
have very small values of $m_d^2$, vanishing for $L \to \infty$). The data for lower $vL^2$ do not collapse but do so when instead graphing versus $vL^3$
in Fig.~\ref{fig:odef}(b), which shows a size independent plateau building up with increasing system size. This plateau reflects the stabilization on the
longer survival time scale $L^3$ of a system-spanning domain wall (in some of the SA instances, as in the periodic systems in Sec.~\ref{sec:sascale}).
Here the plateau value not only reflects the fraction of instances with a domain wall, but also an average over all the possible locations of the domain
wall that determine the value of $m_{dx}$, Eq.~(\ref{mdxdef}), or the similarly defined
$m_{dy}$. In principle some refined domain-wall order parameter could be defined to maximize the signal, e.g., associating values $1$ and $0$ with the
presence and absence, respectively, of a system-spanning domain wall identified using a simple modification of the winding number calculation in
Sec.~\ref{sec:wind}. However, the current definition $m_d^2$ is convincing enough for drawing our conclusions on the $L^3$ time scale of domain walls here.

\section{Quantum Annealing}
\label{sec:QA}

For the purpose of QA, we here write the Hamiltonian Eq.~(\ref{qham}) with time dependent couplings as
\begin{equation}
H(t) = -J(t)\sum_{\langle ij\rangle}\sigma^z_i\sigma^z_j - \Gamma(t)\sum_i\sigma_i^x,
\end{equation}
with the following time dependence:
\begin{equation}\label{jgtime}
J(t) = \left(\frac{t}{t_{\rm max}}\right)^2,\quad \Gamma(t) = \left(1-\frac{t}{t_{\rm max}}\right)^2.
\end{equation}
The quadratic form at $t=0$ is chosen to reduce non-analyticity at the beginning of the protocol. The forms of $J(t)$ and $\Gamma(t)$ are
also reminicent of those in D-Wave devices \cite{johnson10,harris18,king22,king23}, which along with Rydberg atom arrays
\cite{keesling19,scholl20,ebadi22,manovitz25,zhang25} likely present the best experimental platform to which our results can be applied.

The process starts at $t=0$ with the system prepared in the ground state of the transverse field, followed by evolution under the time dependent
Schr\"odinger equation to $t=t_{\rm QA}$. For conventional physical observables, we calculate the sqared magnetization, the total energy density, and
the Ising energy [without the factor $J(t)$] density:
\begin{subequations}
\begin{eqnarray}
	m^{2} & = & \left\langle \psi(t)\left\vert\Bigg( \frac{1}{N}\sum\limits_{i\in N}\sigma^z_{i}\Bigg)^{2}\right\vert\psi(t)\right\rangle,\\
	e & = & \left\langle \psi(t)\left\vert \frac{H(t)}{N}\right\vert\psi(t)\right\rangle,\\
	e_z & = & \left\langle \psi(t)\left\vert\frac{1}{N}\sum_{\langle ij\rangle}\sigma^z_i\sigma^z_j\right\vert\psi(t)\right\rangle.
\end{eqnarray}
\label{meez}
\end{subequations}
We also compute the fidelity, i.e., the probability of the system in the evolved state $|\Psi(t)\rangle$ to be in the instantaneous ground state
$|\Psi_0(t)\rangle$ of H;
\begin{equation}\label{fidelity}
F_{\rm GS}(t) = |\langle \Psi_0(t)|\Psi(t)\rangle|^2.
\end{equation}
In practice it is convenient to analyze the (natural) logarithm with a minus sign, $-{\rm log} F_{\rm GS}$, which is positive and approaches
$0$ as $v \to 0$. We refer to it as the log fidelity.

We study these quantities as functions of the ratio $s(t) \in [0,1]$ between the instantaneous time $t$ and the total time for annealing to the classical limit,
\begin{equation}
s={t}/{t_{\rm max}},
\end{equation}
focusing on the observables at the classical limit $s=1$ of the process and at the point $s=s_{\rm c}$ where the system is critical in the adiabatic limit,
\begin{equation}
s_{\rm c} = \frac{1}{1+\sqrt{(\Gamma/J)_{\rm c}}} \approx 0.3643,
\end{equation}
where we use $(\Gamma/J)_{\rm c}=3.04458$ \cite{liu13}. Because of the small system sizes, the results are insensitive to small deviations from the
true critical point. For scaling analysis, we consider a wide range of annealing velocities, $v=1/t_{\rm QA}$, and system sizes $L \in \{3,4,5,6\}$.

Unlike the classical Ising model discussed in Sec.~\ref{sec:SA}, time evolution the 2D TFIM is a computationally challenging problem and there
are few reliable numerical results. While the corresponding 1D system can be mapped to free fermions and has been studied exactly on large length and time
scales \cite{polkovnikov05,zurek05,dziar05,dziar10,polk11,delcampo14,king22,zeng23,grabarits25}, the 2D model can be studied in real time only with methods
limited to small lattices and/or short times.
Beyond exact calculations for small systems that we report on here (and for which we are not aware of previous comprehensive work), recent research
on the TFIM and other quantum many-body systems have focused on efficient methods based on the time dependent variational principle in the context of
matrix- and tensor-product states \cite{haegeman11,haegeman16,zcarnik19,yang20,dziarmaga21,unfried23,li24,tindall25}, neural-networks \cite{schmitt20},
and other variational wave functions \cite{carleo12,carleo14,blas16,mauron25}. While the above methods represent impressive developments, there are still
severe limitations in reaching long times with maintained fidelity \cite{schmitt22,king25}.

Instead of resorting to more advanced but approximate methods, we here carry out exact time evolution for small lattices but with essentially
no restrictions on the annealing time. For the integration of the Schr\"odinger equation we work with an exact representation of the wave function,
taking advantage of all lattice symmetries (translational and point-group) and the $\mathbb{Z}_2$ spin-reflection symmetry. Thus, we work in the restricted
Hilbert space in the same fully symmetric sector as the initial state with $\sigma^x=1~\forall ~i$, i.e., zero momentum and even wave function with respect to all
reflections and rotations. The number of states in this symmetry block is about $2 \times 10^8$, and extending the calculations to $L=7$ would not
be feasible. We solve the Schr\"odinger equation using an explicit Runge-Kutta method of order $8(5,3)$ with adaptive step size \cite{hairer93}. 
The absolute and relative tolerances for the step size adjustment were set to $10^{-12}$. 

\subsection{Kibble-Zurek Scaling}
\label{sec:QA:nes_crit}

KZ scaling in the TFIM has in the past been accessed using quantum MC methods implementing imaginary-time evolution, for which the exponents are
the same as in real time though the scaling functions are different \cite{degrandi10,degrandi11,degrandi13,liu13}. Critical scaling has also been
recently studied by simulations mixing classical and quantum fluctuations \cite{hotta23}. For the purpose of comparing with QA experiments, true
real-time results are required, however. We here test for real-time KZ scaling in the $L \le 6$ systems at the infinite-size quantum-critical point
$s_c$ and, in the case of the fidelity, also at the final point $s=1$. Other quantities do not exhibit KZ scaling at $s=1$ and we study their different
scaling properties in that limit in Sec.~\ref{sec:QA:nes_end}.

To use the KZ finite-$L$ ansatz Eq.~(\ref{chiv}) for QA ending at the critical point, we need the critical exponents for the different
quantities discussed above. The generic finite-size scaling exponent $\kappa/\nu$ can also be expressed with the scaling dimension $\Delta_A$ of
the observable $A$; $\kappa/\nu=\Delta_A$. The quantum-critical point of the 2D TFIM is in the universality class of the 3D classical Ising model
($d+z$ dimensions with $d=2$ and $z=1$), following from the mapping to $2+1$ space-time dimensions. Very precise values for the scaling dimensions are
available \cite{poland19}. In the case of the full energy density $e$, the scaling dimension is that of the Hamiltonian itself; $\Delta_e=d+z$. The Ising
interaction and the transverse field can both drive the transition without changing the symmetry [and in our QA protocol Eq.~(\ref{jgtime})
they change simultaneously]; thus they have the same scaling dimensions $\Delta_{e_z}=\Delta_{e_x}$, and this value also defines the correlation-length
exponent $\nu$; $1/\nu=d+z-\Delta_{e_z}$, with $\Delta_{e_z}=1.412625$ and, therefore, $1/\nu=1.587375$. For the squared order parameter
$\Delta_{m^2}=2\Delta_m=2\beta/\nu=1.036298$, where $\beta$ is the standard exponent of the order parameter.

\begin{figure}[t]
\includegraphics[width=80mm]{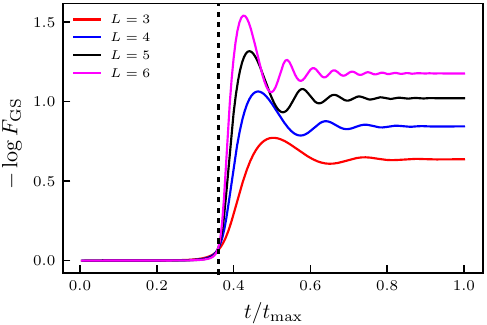}
\vskip-1mm
\caption{Log fidelity vs time during QA for different system sizes. The total annealing time is increased with the system size as $t_{\rm max} = L^2$.
The dashed line indicates the known equilibrium critical point for $L \to \infty$.}
\label{fig:logf_time}
\end{figure}

As a first test of KZ scaling it is useful to consider the fidelity, Eq.~(\ref{fidelity}), the logarithm of which approaches zero in the adiabatic
limit. The time evolution of the log fidelity is shown in Fig.~\ref{fig:logf_time} for all the systems that we have studied. Here the total
annealing time is scaled with the system size as $t_{\rm max}=L^2$ as an example. We have also marked the location of the critical point.
The fidelity indeed starts to decrease sharply as this point is traversed, with the graphed $-\log F_{\rm GS}$ exhibiting a steep increase followed by
a peak of height that increases with $L$. This behavior implies that the QA time $t_{\rm max}$ must increase faster than $L^2$ in order to keep the system
near adiabatic ($F_{\rm GS}$ close to $1$), which, as we will see below, is in accord with KZ scaling. We will here not be concerned
with the oscillations in $-\log F_{\rm GS}$ that take place inside the ordered phase in Fig.~\ref{fig:logf_time}, but note the flattening out to almost
constant as the classical limit is approached for all the system sizes. This behavior reflects the growing gap between the ground state and the
excitations for $\Gamma > \Gamma_c$ (the gap minimum being at $\Gamma_c$ with small finite-size corrections), which allows the excited states to
thermalize among themselves at the instantaneous value of $s(t)$ \cite{blas16,deutsch18,banks19} while the probability of the system being in the
instantaneous ground state is conserved.

\begin{figure}[t]
\includegraphics[width=80mm]{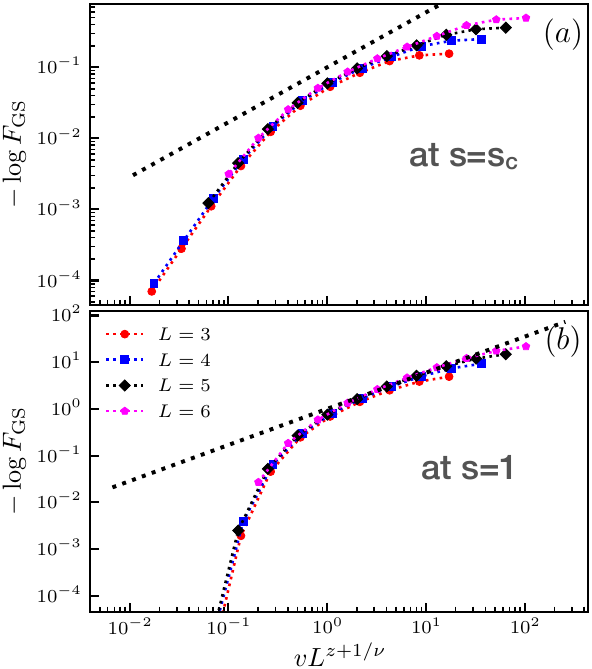}
\vskip-1mm
\caption{Log fidelity at the infinite-size critical point $s_c$ in (a) and in the classical limit $s=1$ in (b).
The annealing velocity $v$ has been scaled with the KZ time $L^{z+1/\nu}$ with $z=1$ and $1/\nu=1.587375$. The lines have the
slopes expected at large $x=vL^{z+1/\nu}$ for a system exhibiting KZ scaling.}
\label{fig:logf_scaled}
\end{figure}

The fidelity is a dimensionless quantity in scaling and we can then test the corresponding KZ ansatz,
\begin{equation}\label{logfkz}
-\log F_{\rm GS} = f_F(vL^{z+1/\nu}),
\end{equation}
by simply graphing $-\log F_{\rm GS}$ versus $x=vL^{z+1/\nu}$, as we do in Fig.~\ref{fig:logf_scaled}(a) for the case of QA to the critical point $s_c$.
There are no adjustable parameters and the data exhibit surprisingly good data collapse, considering the small system sizes. In the limit of small
values of $x$, the scaling function $f_F(x)$ onto which the data collapse takes the form $f_F \sim x^2$. This behavior is expected from adiabatic
perturbation theory (APT) \cite{degrandi10,degrandi11}, which applies in the low-velocity limit of finite systems because of their critical  gaps
$\Delta(L) \propto L^{-z} = L^{-1}$. APT predicts a $v^2$ behavior, multiplied by a power of $L$ compatible with KZ scaling. The fact that
$x \propto v$, implies the observed quadratic behavior in $x$.

In Fig.~\ref{fig:logf_scaled}(a), the data for each system size deviate from the common scaling function in the high-velocity limit, where the
$v \to \infty$ value for given $L$ corresponds to the overlap between the critical state and the trivial initial state. This overlap of course decreases
with increasing $L$, but a KZ scaling regime gradually builds up as $L$ increases, with the high-$v$ crossover shifted to larger values of $x$.
The log fidelity here is expected to be proportional to $L^d$ \cite{degrandi11,kolodrubetz12,degrandi13}, which corresponds to
the ground state probability being exponentially small, $F_{\rm GS} \sim {\rm exp}[{-a(L/\xi_v)^d}]$ with a velocity independent $a$. The scaling form
Eq.~(\ref{logfkz}) then implies $f_F \to (vL^{z+1/\nu})^{d/(z+1/\nu)}$, which is indicated by the dashed line in Fig.~\ref{fig:logf_scaled}(a). Here the
match with the data is not perfect but appears to improve slowly with increasing system size. The plausibility of the slow approach is supported
by the fact that the finite-size critical points, which can be taken as the $t/T_{\rm max}$ point of the maximum value of $-\log F_{\rm GS}$ in data such
as in Fig.~\ref{fig:logf_time}, are still quite far away from the $L \to \infty$ critical point for the small system sizes considered here.

The regime of asymptotic power law should formally correspond to $1 \ll \xi_v \ll L$ and large $L$, which is commonly referred to as the KZ
scaling regime. However, it should again be noted that the KZ time scale $L^{z+1/\nu}$ also applies when $L < \xi_v$, where the QA process for finite $L$
approaches fully adiabatic. The entire regime $\xi_v \gg 1$ is therefore governed by the finite-size KZ mechanism. Despite the finite-size corrections to the
scaling function in Fig.~\ref{fig:logf_scaled}(a), the fact that the data do collapse versus $vL^{z+1/\nu}$ both in the near-adiabatic and KZ regimes is quite
remarkable, given the small system sizes. To improve the data collapse in the KZ scaling regime, the maximum values of $-\log F_{\rm GS}$, the locations of
which can be taken as $s_c(L)$ [and are seen to drift toward $s_c=s_c(\infty)$ Fig.~\ref{fig:logf_time}] can in principle be used instead of the $s=s_c$
values in Fig.~\ref{fig:logf_scaled}(a). This procedure removes some of the subleading finite-size corrections stemming from the drift of $s_c(L)$. Even
more impressive is that the log fidelity at the final point $s=1$, analyzed in Fig.~\ref{fig:logf_scaled}(b), exhibits almost perfect KZ scaling, not just
with excellent data collapse overall but also reproducing the expected asymptotic power-law form with much less finite-size corrections than at $s_c$. Thus,
while there is a steep increase and rather complicated oscillations in $-\log F_{\rm GS}$ close to the quantum phase transitions, as seen in
Fig.~\ref{fig:logf_time}, the behavior once the oscillations have decayed away reflects a very stable probability of the system to remain in the ground
state after traversing the critical point.

The asymptotic behavior for $x \to 0$ in  Fig.~\ref{fig:logf_scaled}(b) is a much faster approach to $0$ than the $x^2$ form at the critical point
in  Fig.~\ref{fig:logf_scaled}(a), thus showing that a large fraction of the defects produced in the neighborhood of the critical point are removed
in the QA process continuing into the ordered phase (though with the size dependence $L^{z+1/\nu}$ of the overall time scale maintained), as is also
reflected in the decay of $-\log F_{\rm GS}$ after the large maximum in Fig.~\ref{fig:logf_time}. The asymptotic $x\to 0$ behavior seems to be exponential
but we have not analyzed the form further.

\begin{figure}[t]
\includegraphics[width=80mm]{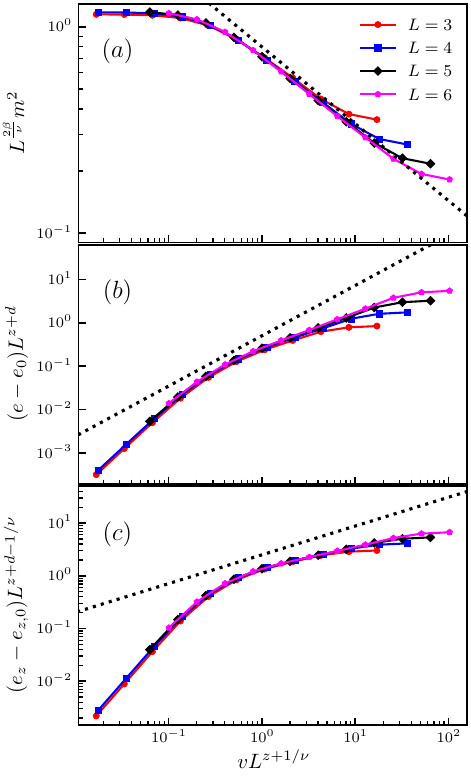}
\vskip-1mm
\caption{Finite-size KZ scaling collapse for the squared order parameter (a), the excess total energy density (b), and the excess Ising energy density
(c). The critical exponents $\nu$ and $\beta$ used in this plot correspond to their known 3D Ising values and $z=1$. The dashed black lines in each of the
panels correspond to the asymptotic KZ power-law behavior for large $vL^{z+1/\nu}$ when $L \to \infty$.}
\label{fig:crit_scaling_2D}
\end{figure}

KZ scaling deep inside the ordered phase is not expected in most observables but is unique to the gap-protected post-critical
ground-state probability (and directly related quantities). As we will see below in Sec.~\ref{sec:QA:nes_end}, observables that depend also on the
properties of the evolved excited states exhibit very different scaling behaviors, originating in coarsening processes similar to those in the classical
model under SA. We therefore do not test for KZ scaling at $s=1$ of the observables defined in Eqs.~(\ref{meez}). As for KZ scaling of these observables
at $s_c$, results are shown in Fig.~\ref{fig:crit_scaling_2D} with vertical scaling of each quantity according to the scaling dimensions discussed
previously. In the case of the energies, we have subtracted their respective size dependent equilibrium values, i.e., we study the excess energies.
Like the log fidelity, in all cases the data collapse is almost perfect without any adjustable parameters but the asymptotic power-law behavior
of the scaling function are again not yet fully manifested, though evolving slowly in the right direction with increasing $L$. In the case of the
squared magnetization in Fig.~\ref{fig:crit_scaling_2D}(a), the asymptotic slope is already almost reproduced.

The asymptotic forms for large $x$ (and large $L$) are again obtained from the KZ scaling ansatz Eq.~(\ref{chiv}) by requiring the trivial $L$ dependence
(with a resulting corresponding $v$ dependence) of the observables in the limit where $\xi_v \ll L$ \cite{liu14}, which in the case of the squared
sublattice magnetization is $m^2 \sim L^{-d}$ while the energy densities are $L$ independent. While the limit $1 \ll \xi_v \ll L$ in which the KZ scaling
function $f(x)$ takes its large-$x$ limiting form of a power law in $x$ may seem formally impossible to realize with system sizes only up to $L=6$, in
practice, as we have seen, we still observe almost the correct KZ power laws in Fig.~\ref{fig:crit_scaling_2D}. The overall KZ time scale $L^{z+1/\nu}$
is also realized there and in the near-adiabatic $L \ll \xi_v$ regime. Our interpretation of this surprisingly good scaling for small systems
is that the subleading finite-size corrections are small in the near-adiabatic regime $L \ll \xi_v$  and the crossover to the KZ power laws is fast (as
manifested in the rather sudden changes in slopes in Figs.~\ref{fig:logf_scaled} and \ref{fig:crit_scaling_2D}). The only minor deviations from the
KZ power laws show that the corrections are small also when $L \gg \xi_v$ is not fully realized.

In the adiabatic limit, $m^2L^{2\beta/\nu}$ approaches a constant at $s_c$, as is clearly seen in Fig.~\ref{fig:crit_scaling_2D}(a); the approach to the
constant form should be analytic, given that the generic scaling function $f(x)$ in the KZ scaling form Eq.~(\ref{chiv}) is analytic in $x$. To analyze
the the adiabatic limits of the excess energies, we have to take into account that the equilibrium finite-size energies have been subtracted off in 
Fig.~\ref{fig:crit_scaling_2D}, while the original generic KZ form Eq.~(\ref{chiv}) (with $\kappa/\nu$ replaced by $\Delta_e$ or $\Delta_{e_z}$)
assumes that the excess is defined with respect to the thermodynamic limit (i.e., the scaling is for the combined
effects of finite size and finite velocity at the critical point). The subtracted finite-size equilibrium energy simply corresponds to removing the
constant term in the scaling function, i.e., with the example of the total energy density,
\begin{subequations}
\begin{eqnarray}
  &&L^{\Delta_e}[e(L,v)-e(L,0)] = f_e(vL^{1/\nu})-f_e(0)~~~ \label{gf1} \\
                            && ~~~= ~g_e(vL^{1/\nu}) \propto (vL^{1/\nu})^a + \ldots,  \label{gf2}
\end{eqnarray}    
\end{subequations}
where we have defined a new scaling function $g_e(x)$ with asymptotic form $x^a$, with the integer valued exponent $a \ge 1$ representing the leading
non-adiabatic corrections. Similarly, we can define $g_{e_z}$ for the excess Ising energy. The question then is what the leading exponents $a$ are for
these excess energies.

As we are dealing with finite systems that always have nonzero instantaneous gaps, APT can again be applied \cite{degrandi10,degrandi11}. When stopping at the
critical point, if the observable commutes with $H(t)$ at the time of measurement, then the leading-order correction to the adiabatic limit should be
quadratic in $v$, which obviously applies to the total excess energy $e-e_0$ (and also the case of the log fidelity analyzed in Sec.~\ref{sec:QA:nes_crit}).
In other cases the dependence is in general linear in $v$. However, the constant of proportionality is the Berry curvature, which vanishes unless the operator (or
the Hamiltonian itself) breaks time-reversal symmetry \cite{degrandi13}. The Ising energy being time-reversal symmetric, we also expect a quadratic $v$ dependence
for $e_z-e_{z,0}$ (unlike the case of imaginary time QA, where the constant is the metric tensor, which does not vanish and the low-$v$ form is linear
\cite{degrandi11,degrandi13}). For both energies, powers of the system size multiply the $v^2$ dependence. Since the KZ scaling functions $g_{e}(x)$ and
$g_{e_{z}}(x)$ must obey the limiting near-adiabatic power laws when $x \to 0$ and $x \propto v$, we should simply have $g_{e}(x) \sim x^2$ and $g_{e_z}(x) \sim x^2$,
both of which are indeed reproduced essentially perfectly by the data in Fig.~\ref{fig:crit_scaling_2D}(b) and Fig.~\ref{fig:crit_scaling_2D}(c).

\subsection{Ordering kinetics and defects}
\label{sec:QA:nes_end}

KZ scaling provides a generic framework for understanding QA ending at a quantum-critical point, and generalizations applying at the initial stages of
entering an ordered phase have also been proposed \cite{schmitt22,king23,samajdar24}. Much less is know in general about the dynamics that ensues when
annealing deep into the ordered phase. It is commonly understood that the excitations created
at the critical point "freeze out" in the ordered phase, in which case KZ scaling would apply all the way to the classical limit of the QA protocol.
Indeed, in cases where all kinetic ordering processes are much slower than the critical fluctuations governing KZ scaling, the final classical state
will to a high degree reflect the critical state. Such a ``memory'' of the critical state has been observed in QA annealing of programmable spin glasses
in D-Wave devices \cite{king23,king25}. Extremely slow dynamics of random in systems corresponding to hard optimization problems (e.g., spin glasses)
should in general be due to a phenomenon similar to Anderson localization that is associated with a high density of exponentially small gaps in the
ordered (or glass-type) phase \cite{altshuler10}. In contrast, as observed in Sec.~\ref{sec:sascale}, and previously with less detail in Ref.~\cite{biroli10},
much faster coarsening dynamics and interface fluctuations take over when crossing into the ordered phase in classical SA of the uniform 2D ferromagnet.
The emergence of similar faster ordering processes beyond the KZ mechanisms should also be expected in QA of uniform quantum system (and perhaps in many
disordered ones as well) \cite{chandran12,gagel15,samajdar24}.

In the case of QA of the Ising model, the most plausible fast ordering mechanism (which is also supported by our study) is thermalization of the gap-protected
excitations in the ordered phase \cite{chandran13a,shimizu18,libal20}, with the essentially conserved excess energy supplied to the system when crossing the
phase transition. However, the lack of reliable numerical methods (and, as our work demonstrates, the neglect so far of small systems) in dimensions higher
than one has inhibited unbiased computational progress on the late-time QA dynamics. Calculations have been largely restricted to mean-field models
\cite{caneva08,wauters17}, field-theory analysis \cite{chandran13a}, large-$N$ approximations \cite{maraga15,maraga16}, and models that map to free
fermions \cite{polkovnikov05,zurek05,dziar05,dziar10,polk11,delcampo14,king22,zeng23,grabarits25,caneva08}. Some of these works have demonstrated domains
formed by coarsening dynamics \cite{chandran13a,maraga16}, while in other works it was argued that the domain sizes are predicted by the Landau-Zener
approximation of annealing \cite{zurek05,caneva08,wauters17}.

\begin{figure}[t]
\includegraphics[width=80mm]{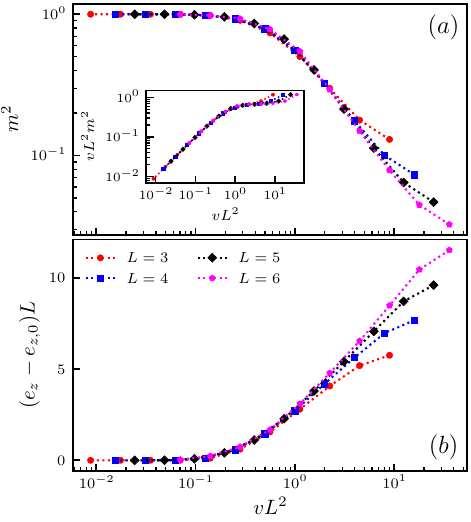}
\vskip-1mm
\caption{Scaling at $s=1$ of the squared order parameter (a) and the excess Ising energy density multiplied by $L$ (b), both vs $vL^2$.
The observed data collapse when scaling the velocity by $L^2$ corresponds to ordering by coarsening dynamics of confined defect
domains.}
\label{fig:end_scaling_2D}
\end{figure}
  
\begin{figure}[t]
\includegraphics[width=80mm]{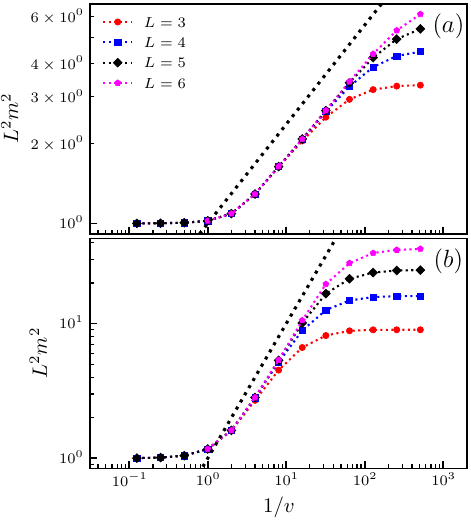}
\vskip-1mm
\caption{The squared magnetization scaled by the system volume, graphed vs the inverse QA velocity so that collapse onto a common scaling function should
take place in the high-velocity limit as well as (a) in the KZ scaling regime for a process measured at $s_c$ and (b) in the coarsening regime when measuring
at $s=1$. The dashed line in (a) shows the slope expected (the line being offset from the data for clarity) in the KZ regime, $f(1/v) \propto (1/v)^{0.4267}$.
In (b) the slope of the dashed line  corresponds to the stage of coarsening when the domains are still of size $l \sim L$, where $f(1/v) \propto 1/v$. In
both cases, there is no collapse to a common scaling function for large $1/v$, where the systems are nearly adiabatic and the data in (a) and (b) instead
collapse when graphing as in Fig.~\ref{fig:crit_scaling_2D}(a) and Fig.~\ref{fig:end_scaling_2D}(a), respectively.}
\label{fig:end_scaling_2D_2}
\end{figure}

Similar to the SA case discussed in Sec.~\ref{sec:sascale}, we are here interested in the types of domains that form at the end of the QA protocol
and their associated time scales of elimination. We will investigate the squared order parameter and the Ising energy density (which at $s=1$ is the
same as the total energy density) and test whether these quantities obey the same type of scaling form as in the SA case, Eqs.~(\ref{scforms}).
We also develop methods to specifically probe the excited states, which allows better access to the rare system-spanning topological defects
that remain at the end of slow QA processes.

To first test for coarsening dynamics with the same exponent $\alpha=2$ as in classical SA, in Fig.~\ref{fig:end_scaling_2D} we plot $m^2$ and
$L(e_z-e_{z,0})$ against $vL^2$, as previously in the case of SA in Figs.~\ref{fig:m2all}(c) and \ref{fig:dE}(b). Both quantities show a very good scaling
collapse, indicating that $\alpha=2$ is most likely the correct power-law relationship between the domain size and time scale in this velocity regime.
In the regime where the velocity-limited correlation length is significantly smaller than the system size, $\xi_v \ll L$ [where now $\xi_v$ does
not follow the KZ form Eq.~(\ref{xiv}) since the QA process is no longer in the vicinity of the critical point], the squared order parameter
must scale as $m^2 \sim 1/L^2$. This regime should correspond to Fig.~\ref{fig:end_scaling_2D}(a) where the collapsed data fall on a straight line on
the log-log plot. In the coarsening form $m^2 = f_m(vL^2)$ the scaling function must have the limiting behavior $f_m(x) \sim 1/x$ for large $x$, provided
that both $x=vL^2$ and $L$ are large enough. The asymptotic form is demonstrated in the inset of Fig.~\ref{fig:end_scaling_2D}(a) with $m^2(x)$ multiplied
by $x$ so that a constant behavior should apply for large $x$. While the data for each $L$ peels off from the approximately flat portion of the almost
collapsed function, the deviations are pushed to larger $x$ with increasing $L$ and the plateau also becomes more flat. 

The size independence of $L^2m^2$ in a disordered system with finite correlation length also of course must apply in the KZ scaling regime with $L \gg \xi_v$ (now with
$\xi_v$ taking the KZ form), which we return to momentarily to contrast the KZ and coarsening behaviors. In Fig.~\ref{fig:end_scaling_2D_2} we graph the
inverse-velocity dependence of $L^2m^2$ both at $s_c$, in Fig.~\ref{fig:end_scaling_2D_2}(a), and at $s=1$ in Fig.~\ref{fig:end_scaling_2D_2}(b), along with
the expected power-law forms. At $s_c$ the exponent $a$ in $L^2m^2 = (1/v)^a$ is obtained from KZ scaling by demanding $m^2L^2=L^{2+2\beta/\nu}(vL^{z+1/\nu})^a$
to be $L$ independent, which gives $a=0.4267$. This KZ scaling regime is manifested in Fig.~\ref{fig:end_scaling_2D_2}(a) versus $1/v$ instead
of $vL^{z+1/\nu}$ used in Fig.~\ref{fig:crit_scaling_2D}(a). Now there is data collapse also in the high-velocity regime (not in the near-adiabatic
low-velocity regime) and the crossover to KZ scaling also happens rather abruptly (like the crossover between KZ and near-adiabatic forms in
Fig.~\ref{fig:crit_scaling_2D}), with only very small corrections. The asymptotic KZ power-law behavior that should be attained for large $1/v$ and
large $L$ is again observed with almost the expected exponent for the largest system and the correct trend of improving with increasing $L$ is clear.
Returning to $s=1$, here we obtain a very different exponent $a=1$ from demanding the $L$ independence of $L^2m^2=L^2(vL^2)^a$. The $1/v$ form is indeed
closely realized in Fig.~\ref{fig:end_scaling_2D_2}(b) and the slope again improves with increasing $L$. While the results in
Figs.~\ref{fig:end_scaling_2D_2}(a) and \ref{fig:end_scaling_2D_2}(b) are not fully converged to their large-$L$ and large-$1/v$ forms, they clearly
support and contrast the respective KZ and coarsening regimes. 

\begin{figure}[t]
\includegraphics[width=75mm, clip]{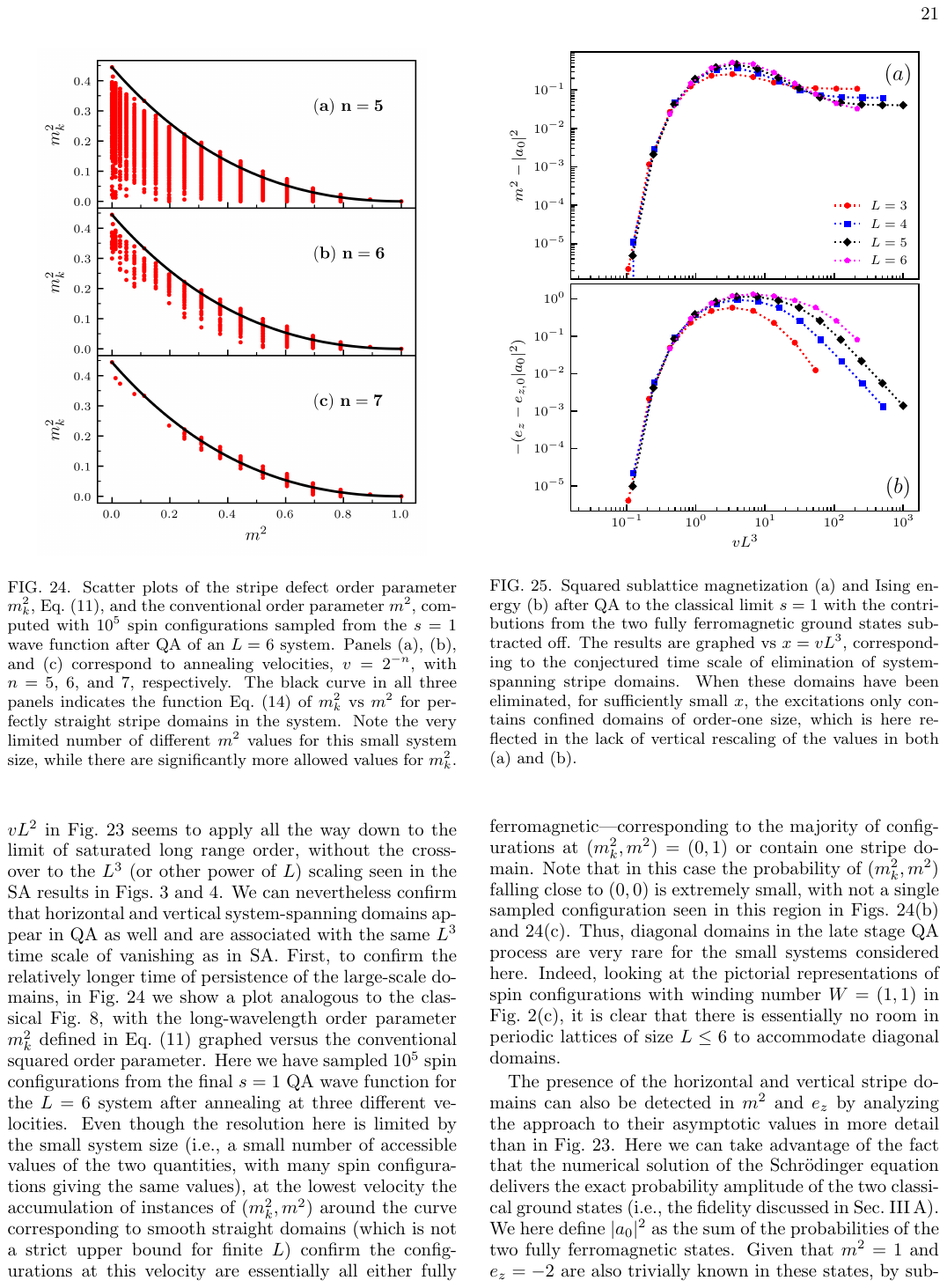}
\vskip-2mm
\caption{Scatter plots in the plane of the conventional order parameter $m^2$ and the stripe defect order parameter $m^{2}_{k}$, Eq.~(\ref{mk}),
based on $10^5$ spin configurations sampled from the $s=1$ wave function after QA of an $L=6$ system. Panels (a), (b), and (c) correspond to annealing
velocities, $v=2^{-n}$, with $n=5$, $6$, and $7$, respectively. The black curve in all three panels indicates the function Eq.~(\ref{mkvsm}) of
$m^{2}_{k}$ vs $m^2$ for perfectly straight stripe domains in the system. Note the very limited number of different $m^2$ values for this small
system size, while there are significantly more allowed values for $m^2_k$.}
\label{fig:mk_v_m2_2D}
\end{figure}

Unlike the case of classical SA, the data collapse versus $vL^2$ in Fig.~\ref{fig:end_scaling_2D} seems to apply all the way down to the limit of
saturated long range order, without the crossover to the $L^3$ (or other power of $L$) scaling seen in the SA results in Figs.~\ref{fig:m2all} and
\ref{fig:dE}. We can nevertheless confirm that horizontal and vertical system-spanning domains are prevalent in late-stage QA as well and are associated
with the same $L^3$ time scale as in SA. We will also show that their effects on the order parameter and excess energy are small on the scales used in
Fig.~\ref{fig:end_scaling_2D} but that they can be observed clearly when analyzing the data in other ways.

First, to confirm the relatively longer time of persistence of the large-scale domains, in Fig.~\ref{fig:mk_v_m2_2D} we show a scatter plot analogous to
the classical Fig.~\ref{fig:corr}, with the long-wavelength domain order parameter $m^2_k$ defined in Eq.~(\ref{mk}) graphed versus the conventional squared
order parameter. Here we have sampled $10^5$ spin configurations from the final $s=1$ QA wave function for the $L=6$ system after annealing at three
different velocities. Even though the resolution here is limited by the small system size (i.e., a small number of accessible values of the two quantities,
with many spin configurations giving the same values), at the lowest velocity the accumulation of instances of $(m^2,m^2_k)$ around the curve Eq.~(\ref{mkvsm})
corresponding to smooth straight domains (which is not a strict upper bound for finite $L$) confirm the configurations at this velocity are
essentially all either fully ferromagnetic, corresponding to the majority of configurations at $(m^2,m^2_k)=(1,0)$, or contain one stripe domain.
Note that, in this case the probability of $(m^2,m^2_k)$ falling close to $(0,0)$ is extremely small (unlike the classical results in Fig.~\ref{fig:corr}),
with not a single sampled configuration seen in this region in Figs.~\ref{fig:mk_v_m2_2D}(b) and \ref{fig:mk_v_m2_2D}(c). Thus, diagonal domains in the
late stage QA process are very rare for the small systems considered here. Indeed, looking at the pictorial representations of spin configurations with
winding number $W=(1,1)$ in Fig.~\ref{fig:config}(c), it is clear that there is essentially no room in periodic lattices of size $L\le 6$ to
accommodate diagonal domains.

\begin{figure}[t]
\includegraphics[width=78mm,clip]{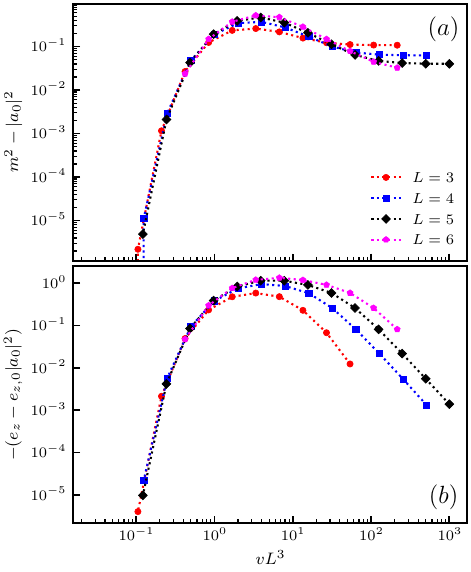}
\vskip-1mm
\caption{Squared sublattice magnetization (a) and Ising energy (b) after QA to the classical limit $s=1$ with the contributions from the two
fully ferromagnetic ground states subtracted off. The results are graphed vs $x=vL^3$, corresponding to the conjectured time scale of elimination
of system-spanning stripe domains. When these domains have been eliminated, for sufficiently small $x$, the remaining excitations only contain
confined domains of order-one size, which is here reflected in the lack of vertical rescaling in both (a) and (b).}
\label{fig:qal3}
\end{figure}

The presence of the horizontal and vertical stripe domains can also be detected in $m^2$ and $e_z$ by analyzing the approach to their asymptotic values in
more detail than in Fig.~\ref{fig:end_scaling_2D_2}. Here we can take advantage of the fact that the numerical solution of the Schr\"odinger equation delivers
the exact probability amplitude of the two classical ground states (i.e., the fidelity discussed in Sec.~\ref{sec:QA:nes_crit}). We here define $|a_0|^2$
as the sum of the probabilities of the two fully ferromagnetic states. Given that $m^2=1$ and $e_z=-2$ are also trivially known in these states, by subtracting
$|a_0|^2$ and $-2|a_0|^2$, respectively, from the $s=1$ values of $m^2$ and $e_z$ we can analyze the contributions stemming specifically
from the excited states. Conjecturing that the time scale of elimination of the stripe domains is $\propto L^3$, as in classical SA according to
the results in Sec.~\ref{sec:sascale}, in Fig.~\ref{fig:qal3} we graph results versus $x=vL^3$. Here there is no further rescaling of the results by
powers of $L$, because there are only small (of order-one size) confined domains left when the stripe domains have been eliminated, The fact that
the data for small $x$ in Fig.~\ref{fig:qal3} indeed collapse onto a very fast decaying function for small $x$ confirm this scenario of a time
$\propto L^3$ required before the remaining defects are all very small.

\begin{figure}[t]
\includegraphics[width=78mm,clip]{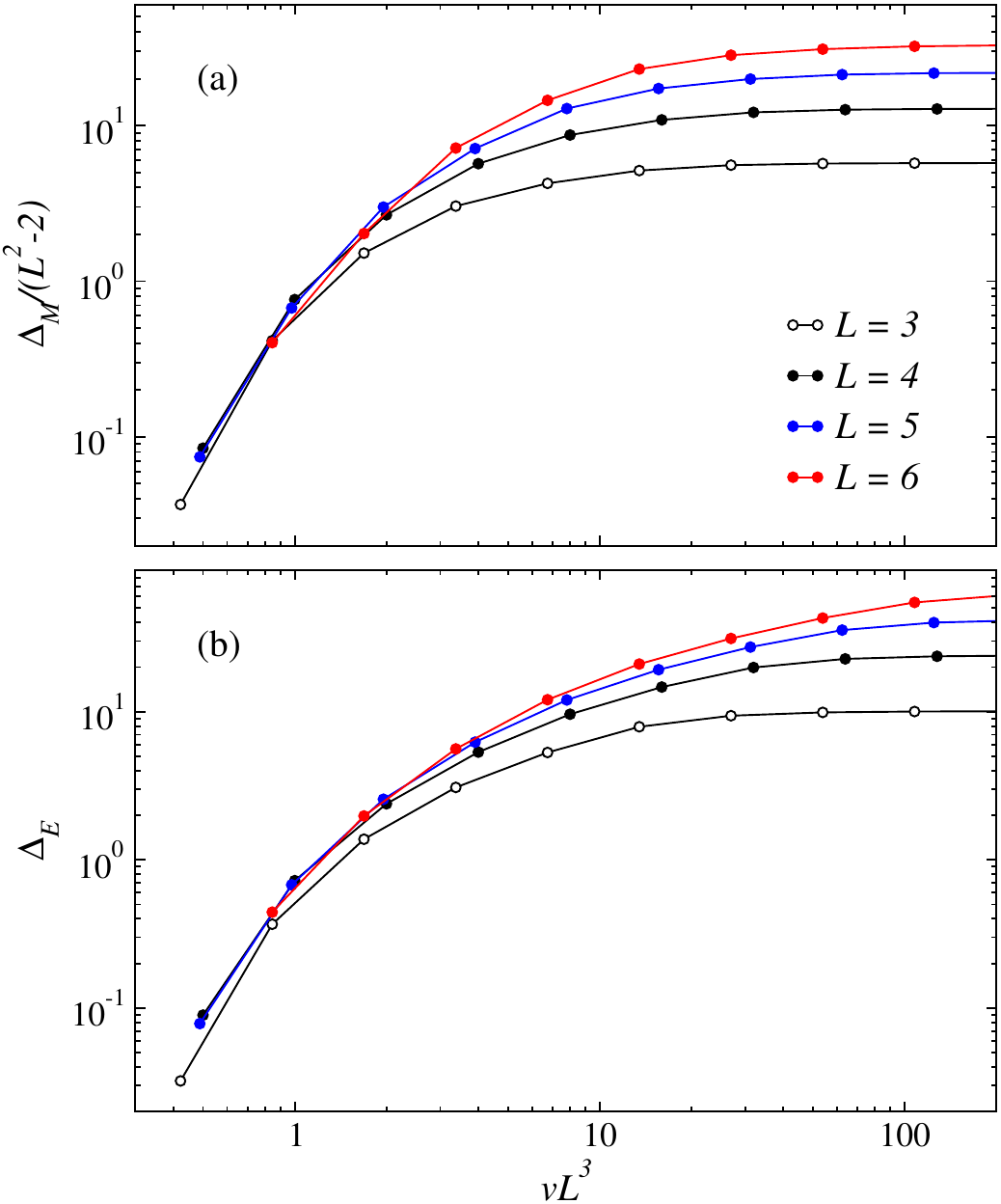}
\caption{The contributions from excited states to (a) the squared magnetization and (b) the total Ising energy, with the corresponding values of the
observables in the lowest excited state subtracted off, as precisely defined in Eqs.~(\ref{deltadefs}). The scaling of the velocity by $L^3$
corresponds to the conjectured time for elimination of stripe domains. The results in (a) have been further rescaled by dividing with the system
volume minus $2$ (the absolute value of $M$ in the lowest excited state), as discussed in the text.}
\label{fig:me3}
\end{figure}

We can further divide the quantities in Fig.~\ref{fig:me3} by the probability $1-|a_0|^2$ of the
system being in any of the excited states. The asymptotic low-velocity forms should then correspond to the properties of the first excited
state of the classical Ising model, which has total energy $E_z=Ne_z=-2N+8$ and total squared magnetization $M^2=N^2m^2=(N-2)^2$. Accordingly,
we define quantities representing positive deviations from these values:
\begin{subequations}
\begin{eqnarray}
\Delta_M & = & (N-2)^2 - \frac{N^2(m^2-|a_0|^2)}{1-|a_0|^2}, \label{deltamdef} \\
\Delta_E & = & \frac{N(e_z+2|a_0|^2)}{1-|a_0|^2} +2N-8 \label{deltaedef}.  
\end{eqnarray}
\label{deltadefs}
\end{subequations}
Results are shown in Fig.~\ref{fig:me3} versus $x=vL^3$. In the case of $\Delta_M$, in order to achieve data collapse for small $x$, the results also have
to be divided by (essentially) the system volume, which is simply explained by the fact that the total magnetization $|M|$ here is a quantity deviating
only by a small amount $\delta$ from $N-2$. With $(N-2-\delta)^2 = (N-2)^2-2(N-2)\delta-\delta^2$, the $(N-2)^2$ term is eliminated in Eq.~(\ref{deltamdef})
and the remaining leading term $(N-2)\delta$ is dealt with by dividing out $N-2$ in Fig.~\ref{fig:me3}(a). The $\delta^2$ term is not important.
The data for both $\Delta_M/(L^2-2)$ and $\Delta_E$, the latter with no rescaling in Fig.~\ref{fig:me3}(b), clearly approach common scaling
functions $f_{M,E}(vL^3)$ for increasing $L$. 

Since the results in Fig.~\ref{fig:me3} represent small differences of large numbers as per Eq.~(\ref{deltadefs}), they are more sensitive to numerical
integration errors than the conventional quantities analyzed previously. These errors become apparent for the longest integration times and (smallest
values of the computed differences) and in Fig.~\ref{fig:me3} we therefore only show results for those velocities for which no numerical anomalies are
apparent. There is still a significant range of points for small $vL^3$ for which the data collapse for $L=4$, $5$, and $6$ is satisfying and clearly
improves with increasing system size. We conclude that the time scale $L^3$ is detectable in QA and that it arises from straight stripe domains,
as evidenced also by the more qualitative results in Fig.~\ref{fig:mk_v_m2_2D}.

\begin{figure}[t]
\includegraphics[width=78mm,clip]{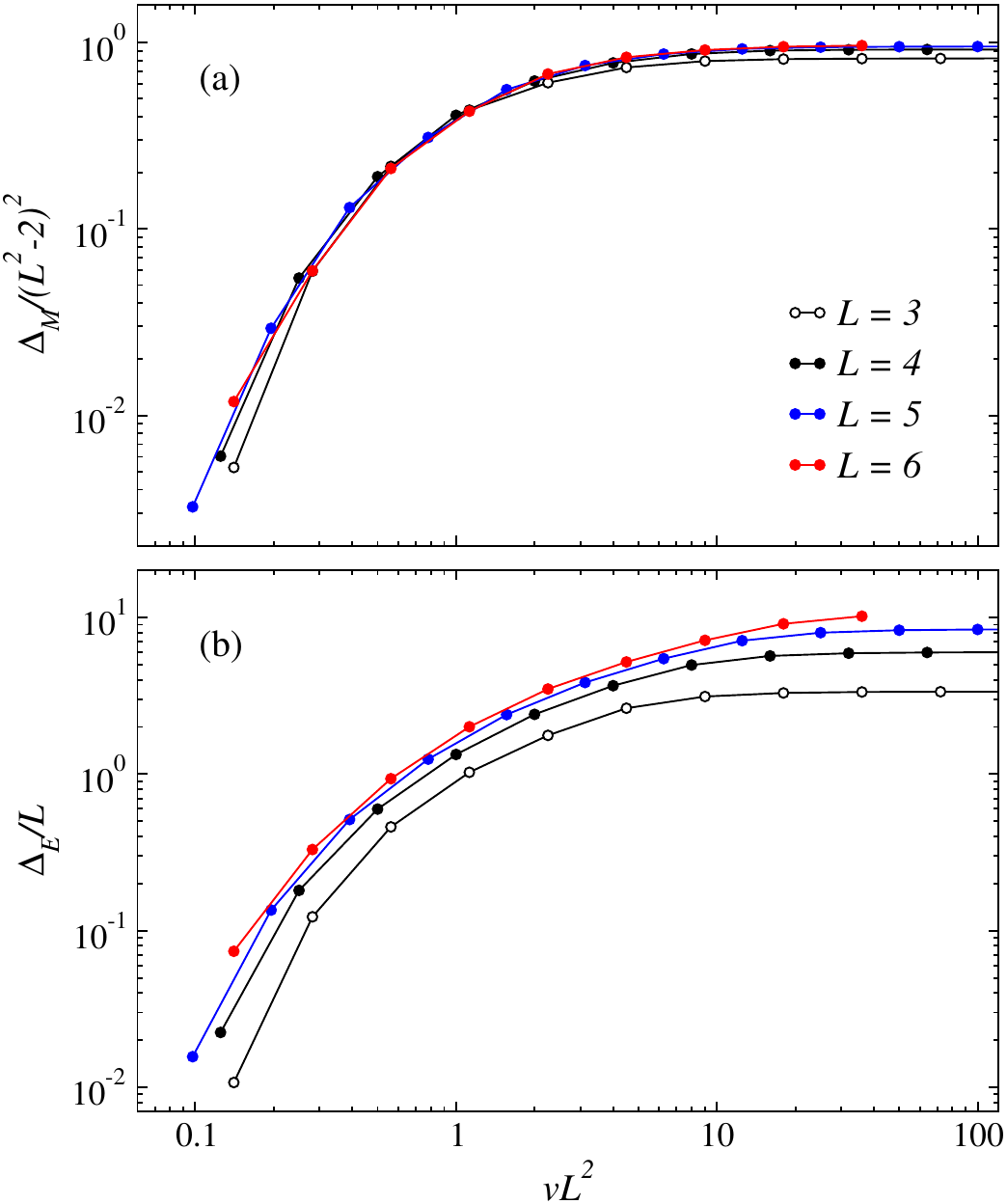}
\caption{The data from Fig.~\ref{fig:me3} graphed versus $vL^2$, with a rescaling of $\Delta_M$ by $1/(L^2-2)^2$ in (a) and of $\Delta_E$ in (b) by $1/L$,
for reasons discussed in the text.}
\label{fig:me2}
\end{figure}

By scaling the same data as in Fig.~\ref{fig:me3} under the assumption that the system still hosts domain walls of length $\propto L$, i.e., at times before
the ultimate convergence of the excitations to those of the lowest energy, we can also recover the conventional coarsening time scale $L^2$, as shown in
Fig.~\ref{fig:me2}. Here the presence of large defect domains, whether confined or system-spanning, corresponds to both terms in Eq.~(\ref{deltamdef})
being of order $N^2$ but not close to canceling each other (in contrast to the other limit considered above). Thus, for scaling we should normalize
$\Delta_M$ by $(N-2)^2$, as we do in Fig.~\ref{fig:me2}(a) when graphing versus $x=vL^2$. For $x \agt 0.5$, the data for $L=4$, $5$, and $6$ collapse
almost perfectly onto a common function, and the tendency to collapse also for small values of $x$ clearly improves with increasing $L$.

In the case of $\Delta_E$, the assumption of remaining large defect domains suggests that $\Delta_E \propto L$ should be analyzed, as we do in
Fig.~\ref{fig:me2}(b). Here the data collapse is not quite as good as in Fig.~\ref{fig:me2}(a), but still clearly improving with increasing $L$. Note
that both quantities in Fig.~\ref{fig:me2} exhibit the poorest data collapse for the smallest $x$ values for all $L$, because those are in the velocity
regime where the large-scale defects have vanished and the results instead show better scaling in Fig.~\ref{fig:me3}. However, even at the
smallest values of $vL^2$, the deviations from the emergent scaling function are not very large, though the same data sets collapse very well in
Fig.~\ref{fig:me3}. Thus, these low-velocity results for small systems are still in a regime where the time scales $L^2$ and $L^3$ cannot be clearly separated
when graphing on a larger scale, as in Fig.~\ref{fig:end_scaling_2D}, thus explaining the lack of visible deviations from conventional coarsening
behavior in said graph. Data for smaller $vL^3$ would be useful in order to extend the range of good data collapse in Fig.~\ref{fig:me3}, and,
conversely, to see larger deviations from a common scaling function for small $vL^2$ and small $L$ in Fig.~\ref{fig:me2}, but this would require
an even more stringent convergence criteria for the demanding long-time numerical solutions of the Schr\"odinger equation.

\section{Conclusions and Outlook}
\label{sec:discuss}

In this paper we first investigated the final-stage $T \to 0$ ordering kinetics in SA of the classical Ising model and then
used insights gained there to analyze the limit of vanishing transverse field in QA of the corresponding quantum model (TFIM) with
a transverse field. The primary difference in focus from previous SA \cite{biroli10} and numerical QA \cite{schmitt22} works on the same model is the role played by
system-spanning topological defects. The original ideas of KZ concerned exactly the slow dynamics of large-scale defects, in the
universe \cite{kibble76} and in liquid $^{\rm 4}$He \cite{zurek85}, but later works on SA and QA have largely neglected them---or, in the
case of QA, involved 1D systems in which the domain walls are point-like and less interesting. Even in the context of classical
phase-ordering kinetics \cite{bray94}, the focus has normally been on the growth of domains by coarsening at finite temperature
in the thermodynamic limit, while we here have studied in detail how both confined and system-spanning minority domains eventually
are eliminated in finite systems under SA and QA.

In SA on periodic lattices (in which domain walls can be classified by the topological winding number), horizontal and vertical stripe defects
vanish on a time scale $L^3$, while diagonal domains have the longer life time $\propto L^{z+1/\nu}\approx L^{3.17}$ corresponding to the KZ
mechanism. With open boundaries, the formal distinction between different types of system-spanning domain walls no loner applies (the winding number
is not defined) and only the $L^3$ scale is manifested for all system spanning defects at the $T=0$ end point of the SA process. The time scales of
elimination of system-spanning stripe defects again exceed the $L^2$ scale of conventional coarsening dynamics, which we also observed in the
open systems using finite-size and -velocity scaling.

The $L^3$ time scale of elimination of system-spanning domain walls in the 2D classical Ising model was found and partially explained some time ago in MC
simulations at fixed temperature $T < T_c$ \cite{lipowski99}, and was further elucidated in the context of sudden quenches to $T=0$ of random states
\cite{spirin01a,spirin01b,olejarz12}. Some of our SA results clearly connect to these previous findings. However, in SA there is the further interesting
and complicating aspect of interplay between the KZ mechanism and ordering kinetics when the system is annealed slowly past a critical point into the
ordered phase. The time scale $L^{3.42}$ of elimination of diagonal domains in periodic systems at fixed $T<T_c$ (where the exponent represents an
improvement over a previous estimate \cite{spirin01a}) is longer than the KZ scale of remaining in equilibrium when traversing the phase transition.
This fact explains why the prevalence of diagonal domains is actually, as mentioned above, controlled by the KZ exponents even for SA all the way
to $T=0$, in clear analogy with the original KZ ideas of ``frozen'' topological defects. However, since the horizontal and vertical domain walls are
eliminated on a shorter time scale in the 2D Ising model below $T_c$, they do not obey KZ scaling in SA into the ordered phase. Moreover, since
diagonal domain walls are not well defined in open systems, the longer time scale associated with them is absent there, and it appears that no
physical observables obey KZ scaling in SA to $T=0$ unless periodic boundary conditions are applied. The important role of the slow dynamics of
system-spanning domain walls carries over to QA as well, where the $L^3$ time scale that we confirm for straight domain walls also in this case
is slower than the KZ scale $L^{2.59}$ and, therefore, the survival probability of these domains should actually obey KZ scaling in QA all the way
to the classical limit (vanishing field), with periodic as well as open boundary conditions.

While our exact numerical QA calculations were limited to very small periodic $L\times L$ lattices with $L \le 6$, the results exhibit
remarkably good scaling. We confirmed KZ scaling of the ground state fidelity, which is protected by the gap in the ordered phase, even in the
extreme limit of QA to vanishing field. Other quantities like the excess energy and the order parameter exhibit KZ scaling
at the critical field, while in the ordered phase the time scale $L^2$ of quantum coarsening is seen very clearly. Though
the time scale $L^3$ of elimination of horizontal and vertical stripe domains is not easily observed in conventional expectation values,
we were nevertheless able to detect this time scale by isolating the contributions from the excited states, which are not subject to KZ
scaling (other than in their overall collective probability) because of their gapless spectrum when $L \to \infty$.

It may appear surprising that systems of size $L \le 6$ can realize all these scaling behaviors---with three different times scales detected---especially
in light of the fact that much larger systems are required to achieve good scaling in SA. It should be kept in mind, however, that the quantum system
is effectively 2+1 dimensional when mapped to a classical problem, and it should therefore be more appropriate to judge the size of the system in terms
of its space-time volume, where the velocity-limited correlation length $\xi_v$ should correspond to the length in the time direction. In the related
problem of eigenstate thermalization, small systems were also sufficient for drawing conclusions about the thermodynamic
limit \cite{rigol08,mondaini16,mondaini17}.

All our QA results in the ordered phase point to the same ordering mechanisms as in SA, though with the twist that the KZ exponents are different
and this affects they way in which the different times scales of ordering kinetics can be observed. Coarsening dynamics of classical systems has also
previously motivated proposals for emergent classical hydrodynamic descriptions of quantum systems obeying eigenstate thermalization
hypothesis \cite{blas16,deutsch18,banks19}. The main idea here is that energy is pumped into the system when it crosses the phase transition,
with subsequent relaxation of the excitations to a finite-temperature state (in which the ground-state probability does not obey Gibbs statistics
because it is essentially conserved after traversing the critical point) for a sufficiently slow QA process. If the resulting temperature is lower
than the critical temperature, the system will equilibrate and begin to form larger domain walls (coarsen), while at higher temperature the system
remains disordered \cite{chandran12,chandran13a}.

This idea of emergent classical hydrodynamics by thermalization of excitations has been checked in various semi-classical works on sudden quenches
\cite{chandran13b,maraga15} and ramps \cite{maraga16}, but, to our knowledge, exact numerical results like those obtained here for a fully quantum mechanical
2D system have not been presented before. The closest case is Ref.~\cite{schmitt22}, which used sophisticated many-body methods for larger system sizes
and found signs of coarsening dynamics. However, due to limitations on the annealing time it was not possible to continue the QA process deep into the
ordered phase and observe the $L^2$ time scale. The fate of the extended domain walls that we have emphasized here was not addressed. A quantum MC method
based on the time dependent variational principle \cite{carleo12,carleo14} was previously applied \cite{blas16} to the same TFIM, but the focus there was
on thermalization and the specifics of critical and ordering dynamics were not addressed. The fact that we also observe the same $L^3$ time scale as in
SA for elimination of system-spanning defects is a further indication of thermalization and emergent classical ordering kinetics in QA.

While our QA study of the TFIM gives a fairly complete picture of the different types of defects and their associated time scales in critical and
ordering dynamics, there are still remaining issues beyond the reach of the exact solutions of small systems employed here. In particular,
it would be interesting to study both straight and diagonal stripe domains more completely in QA, using the winding number methods that we employed
in the context of classical SA. An interesting prospect is that experimental QA platforms may be superior to classical calculations in this regard.
Both D-Wave quantum annealers and Rydberg atom arrays should be suitable for this purpose. In the former, measurements can currently be performed only
in the final annealed state at $\Gamma=0$ \cite{king23,king25}. This limitation is no disadvantage in the present context, which concerns
exactly the final QA stage. It is also not necessary here to strive for very large system sizes, given that already $L \le 6$ results show good scaling.
It would be very interesting to experimentally test the $L^3$ time scale of survival of horizontal and vertical stripe domains in larger systems, as
we have done here already with the small lattices in various ways. It is noteworthy that this time scale is also present in systems with open boundary
conditions, which are easier to implement in practice. Analyzing the surviving system-spanning domains in periodic systems of size
$L=10 \sim 20$ would likely also resolve the outstanding issue of the time-scale of diagonal stripe domains. Currently, fully periodic lattices of native
qubits can be studied with D-Wave annealers for $L \le 12$ \cite{sathe25}. D-Wave devices may also have an advantage here in terms of their short annealing
times possible with the relatively large energy scales of the coupled qubits and efficient state preparation and measurements. This aspect is important,
as a large number of repetitions of the QA process will be required in order to collect sufficient statistics.

Our main specific prediction for QA of the TFIM to $\Gamma=0$ are: (i) The probability $P_{\rm GS}$ of reaching one of the fully ferromagnetic ground states should
obey KZ scaling, i.e., $P_{\rm GS} = f(vL^{z+1/\nu})$ with $z+1/\nu \approx 2.59$. (ii) The probability of observing any system-spanning
domain wall should also obey KZ scaling. (iii) The longer time scales of order $L^3$ (for horizontal and vertical domain walls in both periodic and open systems) and
$L^{3.42}$ (conjectured for diagonal domains in periodic systems) should be manifested in statistical measurements including the excited states only, i.e., all
instances of QA which do not lead to one of the fully ferromagnetic ground states. Sufficiently long annealing times will be required in order to
eliminate most confined defects by coarsening and to establish non-fractal domain walls; otherwise the conventional coarsening processes will
dominate the measurements. The longer time scale of the diagonal domains in periodic systems will also be more difficult to detect because the
horizontal and vertical domains are more prevalent.

While the best way to detect the time scales of system-spanning defects should be to count them explicitly in the measured QA configurations, conventional physical
observables, like the order parameter and the energy should also be useful in this regard. In experiments, when averaged over the classical excited states
only, Eqs.~(\ref{deltadefs2}) for the differences with respect to the lowest excited states can be written more conveniently as just
\begin{subequations}
\begin{eqnarray}
\Delta_M & = & (N-2)^2 - M^2 \label{deltamdef2} \\
\Delta_E & = & E_z +2N-8 \label{deltaedef2},
\end{eqnarray}
\label{deltadefs2}
\end{subequations}
where $M$ is the total magnetization (not divided by the volume) and $E_z$ the total Ising energy. Here Eq.~(\ref{deltaedef2}) applies for periodic
boundary conditions, while for open systems the energy cost $8$ of the smallest local defect should be replaced by $4$, i.e., the energy of a corner
defect. Similarly, for cylindrical boundary conditions the lowest excitation energy, that of a flipped spin on one of the open edges, is $6$. The quantities
$\Delta_M$ and $\Delta_E$ should be analyzed as in Fig.~\ref{fig:me3} to detect the both the times scales $L^3$ and $L^{3.42}$ (which should multiply $v$ on
the $x$ axis) and as in Fig.~\ref{fig:me2} to detect the coarsening time $\propto L^2$.

Beyond the prospect of using QA devices to resolve outstanding issues for one of the prototypical quantum many-body systems, the presence of three different
times scales in QA of the TFIM also provides an opportunity to define a rigorous experimental test-bed. In particular, break-down of scaling at long
annealing times is expected due to finite temperature and, possibly, other noise effects \cite{dutta16,weinberg20,bando21,king22,king23}. It will be useful
to observe the exact manifestations of these imperfections in the different types of critical scaling and defect elimination that we have discussed
here. For small systems, minor randomness in the couplings will likely be less important (while eventually becoming important for large
systems). The 1D TFIM also is an excellent model for such tests \cite{king22}, but the 2D case has much richer scaling behavior, as we have seen,
with clear differences between KZ scaling and ordering kinetics.

It is not clear to us how capable current devices would be to test our predictions (i)--(iii) above, but experiments on the 2D ferromagnet with a previous
generation of the D-Wave device \cite{weinberg20}, as well as more recent D-Wave experiments on 1D systems \cite{king22} and spin glasses \cite{king23,king25},
suggest that at least the coarsening scale $L^2$ and the KZ scale $L^{2.59}$ should be
detectable at the time scales before the noise effects become important with this platform. Recent D-Wave experiments with other models implemented have
already observed some aspects of coarsening \cite{ali24}, as have experiments with Rydberg atoms \cite{manovitz25}. The scaling approaches we have
developed here should help to more clearly observe the time scale of coarsening with current and future devices, and it will of course be particularly 
interesting to test our predictions for system-spanning domain walls. 

The results presented here also pose a constructive challenge for recently improved tensor-network \cite{tindall25} and neural-network \cite{mauron25} methods,
which were claimed to simulate even complex 3D spin-glass QA problems more efficiently than in recent work \cite{king23,king25} with D-Wave annealers. These
claims have themselves been countered \cite{king25b}, with the main point of contention being whether the classical calculations for small system sizes
(tens of qubits) and short times can really be extended with maintained fidelity to systems orders of magnitude larger (hundreds to thousands of qubits
in the experiments) and much longer annealing times. The uniform ferromagnetic 2D TFIM might appear to be a much easier problem for the new algorithms,
but it may still be challenging to reach the long time scales on which system-spanning defects are formed when crossing the phase transition and 
subsequently decay away in the ordered phase.

\begin{acknowledgments}
We would like to thank Mohammad Amin, Anushya Chandran, Leticia Cugliandolo, Andrew King, Paul Krapivsky, Pranay Patel, Anatoli Polkovnikov,
and Jack Raymond or interesting and helpful discussions. This work was supported by the Simons Foundation under Grant No.~511064. The calculations
were carried out on the Shared Computing Cluster managed by Boston University's Research Computing Services. The numerical solutions of the
Schr\"odinger equation were obtained using the QuSpin library \cite{weinberg17}.
\end{acknowledgments}




\end{document}